\documentclass{elsart}
\usepackage{color,setspace,times}
\usepackage{amssymb,amsmath,graphicx,color,rotating,subfigure,url,multirow}
\usepackage{lineno}
\usepackage[square,sort&compress,comma]{natbib}

\journal{Physica A} 

\begin{document}

\begin{frontmatter}

\title{Scaling in the distribution of intertrade durations of Chinese stocks}
\author[BS,SS]{Zhi-Qiang Jiang},
\author[SZSC]{Wei Chen},
\author[BS,SS,RCE,RCSE]{Wei-Xing Zhou\corauthref{cor}}
\corauth[cor]{Corresponding author. Address: 130 Meilong Road, P.O.
Box 114, School of Business, East China University of Science and
Technology, Shanghai 200237, China, Phone: +86 21 64253634, Fax: +86
21 64253152.}
\ead{wxzhou@ecust.edu.cn} %

\address[BS]{School of Business, East China University of Science and Technology, Shanghai 200237, China}
\address[SS]{School of Science, East China University of Science and Technology, Shanghai 200237, China}
\address[SZSC]{Shenzhen Stock Exchange, 5045 Shennan East Road, Shenzhen 518010, China}
\address[RCE]{Research Center for Econophysics, East China University of Science and Technology, Shanghai 200237, China}
\address[RCSE]{Research Center of Systems Engineering, East China University of Science and Technology, Shanghai 200237, China}

\begin{abstract}
The distribution of intertrade durations, defined as the waiting
times between two consecutive transactions, is investigated based
upon the limit order book data of 23 liquid Chinese stocks listed on
the Shenzhen Stock Exchange in the whole year 2003. A scaling
pattern is observed in the distributions of intertrade durations,
where the empirical density functions of the normalized intertrade
durations of all 23 stocks collapse onto a single curve. The scaling
pattern is also observed in the intertrade duration distributions
for filled and partially filled trades and in the conditional
distributions. The ensemble distributions for all stocks are modeled
by the Weibull and the Tsallis $q$-exponential distributions.
Maximum likelihood estimation shows that the Weibull distribution
outperforms the $q$-exponential for not-too-large intertrade
durations which account for more than 98.5\% of the data.
Alternatively, nonlinear least-squares estimation selects the
$q$-exponential as a better model, in which the optimization is
conducted on the distance between empirical and theoretical values
of the logarithmic probability densities. The distribution of
intertrade durations is Weibull followed by a power-law tail with an
asymptotic tail exponent close to 3.
\end{abstract}


\end{frontmatter}

\section{Introduction}

Intertrade duration is the waiting time between consecutive
transactions of an equity, which contains information contents of
trading activity and has crucial relevance to the microstructure
theory
\cite{Diamond-Verrecchia-1987-JFE,Easley-OHara-1992-JF,Dufour-Engle-2000-JF}.
The intertrade duration is a measure of trade intensity, which is
associated with price adjustments. Short intertrade durations lead
to large price change and long intertrade durations minimize the
price impact in a transaction \cite{Jasiak-1999-Finance}. The order
placement process in stock market can be treated as a point process.
Based on estimating event intensity of the marked point process, an
autoregressive conditional duration (ACD) model is proposed by Engle
and Russell for modeling trade duration with temporal correlation
and other financial variables
\cite{Engle-Russell-1998-Em,Engle-2000-Em}. In this framework, the
intertrade duration $\tau$ is modeled as follows
\begin{equation}
\tau = \psi\epsilon~,
\end{equation}
where $\psi$ is the expected value of $\tau$ and $\epsilon$ is an
independent and identically distributed variable, known as the
normalized duration. There are variants of the ACD model, such as
the fractional integrated ACD model \cite{Jasiak-1999-Finance}, the
logarithmic ACD model \cite{Bauwens-Giot-2000-AES}, the threshold
autoregressive conditional duration (TACD) model
\cite{Zhang-Russell-Tsay-2001-JEm}, the stochastic conditional
duration (SCD) model \cite{Bauwens-Veredas-2004-JEm}, and the
stochastic volatility duration (SVD) model
\cite{Ghysels-Gourieroux-Jasiak-2004-JEm}. In the model
specification, there are different assumptions for the distribution
of normalized duration, such as exponential, Weibull, generalized
Gamma, Burr \cite{Sun-Rachev-Fabozzi-Kalev-2008-AF}. We note that
the Burr XII distribution is a general form of the Tsallis
$q$-exponential distribution
\cite{Burr-1942-AMS,Tsallis-1988-JSP,Nadarajah-Kotz-2006-PLA}.

Alternatively, in the econophysics community, the continuous-time
random walk (CTRW) formulism has been adopted to deal with the
intertrade durations and price dynamics
\cite{Scalas-Gorenflo-Mainardi-2000-PA,Mainardi-Raberto-Gorenflo-Scalas-2000-PA,Masoliver-Montero-Weiss-2003-PRE,Scalas-2006-PA,Masoliver-Montero-Perello-Weiss-2006-JEBO}.
By analogy with diffusion, the variations of log prices (or
zero-mean returns) and the associated time intervals are considered
as jump steps and waiting times between consecutive steps in CTRW
processes. According to the CTRW formulism, the probability density
function $p(X,t)$ can be expressed as follows,
\begin{equation}
p(X,t) = \delta(X) C(t) + \int^t_0 {\rm
d}t'f(t-t')\int^{+\infty}_{-\infty}{\rm d}x'\lambda(X-X')p(X',t')~,
  \label{Eq:pdfpxt}
\end{equation}
where $X$ represents the log prices or zero-mean returns, $C(t)$ is
the complementary cumulative function (or survival probability
function), and $f(t)$ is the probability density function. As shown
in Eq.~(\ref{Eq:pdfpxt}), finding $C(t)$ is a key step for obtaining
the expression of $p(X,t)$. By assuming that jump steps and waiting
times are uncorrelated, the complementary cumulative distribution
can be theoretically approximated by the Mattag-Leffler function
\cite{Scalas-Gorenflo-Mainardi-2000-PA,Mainardi-Raberto-Gorenflo-Scalas-2000-PA},
\begin{equation}
C(\tau) = E_{\beta}[-(\tau/\tau_0)^{\beta}] :=
\sum_{n=0}^{\infty}(-1)^n\frac{[-(\tau/\tau_0)^{\beta}
]^n}{\Gamma(\beta n+1)}, ~~\beta > 0.
 \label{Eq:MFfunction}
\end{equation}
Note that Eq.~(\ref{Eq:MFfunction}) has already been verified in
some empirical studies, such as the waiting time distribution of
BUND futures traded on the LIFFE
\cite{Mainardi-Raberto-Gorenflo-Scalas-2000-PA} and 10 stocks traded
in the Irish stock market
\cite{Sabatelli-Keating-Dudley-Richmond-2002-EPJB}. If $\beta = 1$
in Eq.~(\ref{Eq:MFfunction}), the Mattag-Leffler function reduces to
a simple exponential function. If $0 < \beta < 1$, it bridges a
stretched exponential and a power law
\cite{Raberto-Scalas-Mainardi-2002-PA}
\begin{equation} \label{Eq:MF2}
E_{\beta} [-(\tau/\tau_0)^{\beta}] \sim
\left\{ \begin{aligned}
          \exp[-(\tau/\tau_0)^{\beta}/\Gamma(1+\beta)],~~&~~\tau/\tau_0 \rightarrow 0^+,   \\
         (\tau/\tau_0)^{-\beta}/\Gamma(1-\beta),~~&~~\tau/\tau_0\rightarrow \infty.
        \end{aligned} \right.
\end{equation}

Empirical analysis of waiting times for different financial data
unveils that the probability distribution can also be described by
power laws
\cite{Sabatelli-Keating-Dudley-Richmond-2002-EPJB,Yoon-Choi-Lee-Yum-Kim-2006-PA},
modified power laws
\cite{Masoliver-Montero-Weiss-2003-PRE,Masoliver-Montero-Perello-Weiss-2006-JEBO},
stretched exponentials (or Weibulls)
\cite{Bartiromo-2004-PRE,Raberto-Scalas-Mainardi-2002-PA,Ivanov-Yuen-Podobnik-Lee-2004-PRE,Sazuka-2007-PA},
stretched exponentials followed by power laws
\cite{Kim-Yoon-2003-Fractals,Kim-Yoon-Kim-Lee-Scalas-2007-JKPS}, to
list a few. In a recent paper, by rejecting the hypothesis that the
waiting time distributions are described by an exponential
\cite{Scalas-Gorenflo-Luckock-Mainardi-Mantelli-Raberto-2004-QF,Scalas-Gorenflo-Luckock-Mainardi-Mantelli-Raberto-2005-FL}
or a power law \cite{Poloti-Scalas-2008-PA}, Politi and Scalas
reported that both Weibull distribution and $q$-exponential
distribution can be utilized to approximate the intertrade duration
distribution \cite{Poloti-Scalas-2008-PA}. In their empirical
investigation, Politi and Scalas found that the $q$-exponential
compares well to the Weibull. They also argued that the distribution
differing from an exponential is the consequence of the varying
transaction activities during the trading period
\cite{Scalas-Gorenflo-Luckock-Mainardi-Mantelli-Raberto-2004-QF,Poloti-Scalas-2008-PA,Scalas-Gorenflo-Luckock-Mainardi-Mantelli-Raberto-2005-FL,Scalas-Kaizoji-Kirchler-Huber-Tedeschi-2006-PA,Politi-Scalas-2007-PA}.
All this empirical evidence shows that the transaction dynamics can
not be modeled by the Poisson process, rejecting the exponential
distribution of intertrade durations.

An interesting feature can be observed in the empirical results of
Politi and Scalas \cite{Poloti-Scalas-2008-PA}. They have
investigated the Trade and Quotes data set of 30 stocks,
constituents of the DJIA index, traded on the NYSE in October 1999.
The estimated parameters of the Weibull and $q$-exponential seem
close across different stocks. This observation suggests the
possible presence of scaling in the intertrade duration
distributions of different stocks. Such common scaling pattern was
reported for the 30 DJIA stocks over a period of four years from
January 1993 till December 1996
\cite{Ivanov-Yuen-Podobnik-Lee-2004-PRE}. The scaling behavior
underpins common underlying dynamics between different stocks and
has important implications for stock modeling. In contrast, Eisler
and Kert{\'{e}}sz analyzed about 4000 stocks and found that the
scaling was not convincing \cite{Eisler-Kertesz-2006-EPJB}.
Specifically, they divided the stocks into two groups based on the
average intertrade duration of individual stocks and performed an
extended self-similarity (ESS) analysis for each group of data,
which resulted in a nonlinear ESS exponent function with respect to
the order. This conclusion seems quite sound since a marked
discrepancy between the tails can be observed in both studies
\cite{Ivanov-Yuen-Podobnik-Lee-2004-PRE,Eisler-Kertesz-2006-EPJB}.
In this work, we shall show that such scaling in intertrade duration
distributions holds for Chinese stocks and the functional form of
the distribution can also be modeled by the Weibull, as reported in
the DJIA stocks \cite{Ivanov-Yuen-Podobnik-Lee-2004-PRE}. No obvious
discrepancy between the duration distributions is observed in the
Chinese stocks, in contrast to the USA stocks.

This paper is organized as follows. In Section \ref{S1:Data}, we
briefly describe the data sets analyzed. Section \ref{S1:UncondPDF}
investigates the scaling behavior of the unconditional distribution
of intertrade durations. We will fit the corresponding distribution
by means of Weibull distribution and $q$-exponential distribution.
Section \ref{S1:CondPDF} studies the conditional distribution of
intertrade durations. Section \ref{S1:Conclusion} concludes.

\section{Data sets}
\label{S1:Data}

The study is based on the data of the limit-order books of 23 liquid
Chinese stocks listed on the Shenzhen Stock Exchange (SZSE) in the
whole year 2003. The limit-order book records ultra-high-frequency
data whose time stamps are accurate to 0.01 second including details
of every event (order placement and cancelation). There were three
time periods on each trading day in the SZSE before July 1, 2006,
named the open call action (9:15 am to 9:25 am), the cooling period
(9:25 am to 9:30 am), and the continuous double auction (9:30 am to
11:30 am and 13:00 pm to 15:00 pm). During the period of open call
auction, all the incoming orders are executed based on the maximal
transaction volume principle at 9:25 am. In the cooling period, all
orders are allowed to add into limit-order book, but no one is
executed. Only in the period of continuous auction, transaction
occurs based on one by one matching of incoming effective market
orders and limit orders waiting on the limit-order book. Therefore,
only trades during the continuous double auction are considered. In
addition, we stress that no intertrade duration is calculated
between two trades overnight or crossing the noon closing. Assuming
that there are $n$ trades at times $\{t_i: i = 1, 2, \cdots, n\}$
during the time interval from 9:30 am to 11:30 am or from 13:00 pm
to 15:00 pm on a trading day, we obtain $n-1$ intertrade durations
$\tau_i = t_{i+1} -t_i$ with $i=1,2,\cdots,n-1$.

In 2003, only limit orders are allowed for submission in the Chinese
stock market and the tick size of all stocks is one cent of Renminbi
(the Chinese currency). Those orders resulting in immediate
execution, whose prices are equal to or higher than the best ask
price for buys and equal to or lower than the best bid price for
sells, can be called as effective market orders. Effective market
orders can be classified into two types based on their
aggressiveness, namely filled effective market orders and partially
filled effective market orders. Partially filled orders are more
aggressive than filled orders since the former have much larger
price impact \cite{Zhou-2007-XXX}. The trades can be classified into
two types accordingly, called filled trades and partially filled
trades for brevity that are initiated respectively by filled and
partially filled effective market orders. We will consider two
additional classes of durations, involving $\tau_{\rm{F}}$ between
two consecutive filled trades and $\tau_{\rm{PF}}$ between two
successive partially filled trades. Table~\ref{Tb:DataDetail}
provides the number of trades and average intertrade duration for
the 23 Chinese stocks under investigation for all trades, filled
trades, and partial filled trades. We find that $\langle
\tau_{\rm{F}} \rangle \ll \langle \tau_{\rm{PF}} \rangle$ and
$N_{\rm{PF}}\ll N_{\rm{F}}$, since $N_{\rm{F}}
\langle\tau_{\rm{F}}\rangle \approx N_{\rm{PF}}\langle
\tau_{\rm{PF}} \rangle$. Although the time resolution of our data is
as precise as 0.01 second, there are still trades stamped with the
same time. In Table \ref{Tb:DataDetail}, we also presents the
numbers of vanishing intertrade durations for the three types of
trades. We see that their proportions are relatively low. More
information about the SZSE trading rules and the data sets can be
found in
Refs.~\cite{Gu-Chen-Zhou-2007-EPJB,Gu-Chen-Zhou-2008a-PA,Gu-Chen-Zhou-2008b-PA}.

\begin{table}[htb]
\begin{center}
\caption{\label{Tb:DataDetail} The number of trades, the number of
vanishing intertrade durations  and average intertrade duration for
the 23 Chinese stocks}
\medskip
\begin{tabular}{ccccccccccccc}
  \hline\hline
  \multirow{3}*[3.6mm]{Code}&& \multicolumn{3}{c}{All trades}&&\multicolumn{3}{c}{Filled trades}&&\multicolumn{3}{c}{Partly filled trades}\\  %
  \cline{3-5}  \cline{7-9}  \cline{11-13}
   && $N$ & $N^0$ & $\langle \tau \rangle$ && $N_{\rm{F}}$ & $N^0_{\rm{F}}$ & $\langle \tau_{\rm{F}} \rangle$ && $N_{\rm{PF}}$ & $N^0_{\rm{PF}}$ & $\langle \tau_{\rm{PF}} \rangle$ \\\hline
  000001 && 889369 & 17809 & 3.81 && 823603 & 16184 & 4.11 && 65766 & 69 & 51.27 \\
  000002 && 509361 & 10922 & 6.69 && 476457 & 10380 & 7.14 && 32904 & 20 & 101.86 \\
  000009 && 447968 & 5628 & 7.69 && 413531 & 5185 & 8.33 && 34437 & 17 & 98.91 \\
  000012 && 290381 & 2092 & 11.69 && 250078 & 1763 & 13.56 && 40303 & 24 & 83.39 \\
  000016 && 188613 & 1513 & 18.14 && 159497 & 1324 & 21.43 && 29116 & 10 & 116.32 \\
  000021 && 411642 & 4468 & 8.33 && 360503 & 3972 & 9.50 && 51139 & 26 & 66.63 \\
  000024 && 133587 & 1815 & 25.13 && 111691 & 1675 & 30.00 && 21896 & 6 & 150.81 \\
  000027 && 313898 & 8908 & 10.83 && 288205 & 8457 & 11.79 && 25693 & 13 & 128.31 \\
  000063 && 265479 & 9754 & 12.76 && 237957 & 9380 & 14.22 && 27522 & 8 & 120.42 \\
  000066 && 277654 & 2329 & 12.32 && 240016 & 2063 & 14.24 && 37638 & 15 & 89.92 \\
  000088 && 97196 & 7060 & 34.87 && 84183 & 6856 & 40.15 && 13013 & 12 & 250.16 \\
  000089 && 189118 & 5097 & 17.88 && 168599 & 4728 & 20.04 && 20519 & 9 & 160.85 \\
  000406 && 271390 & 3181 & 12.66 && 237390 & 2880 & 14.46 && 34000 & 11 & 100.36 \\
  000429 && 117425 & 564 & 28.79 && 101329 & 496 & 33.30 && 16096 & 3 & 204.91 \\
  000488 && 120098 & 1294 & 28.32 && 95015 & 1158 & 35.76 && 25083 & 5 & 134.35 \\
  000539 && 114722 & 15245 & 29.27 && 98296 & 14812 & 34.10 && 16426 & 15 & 199.73 \\
  000541 && 68738 & 666 & 49.35 && 56232 & 599 & 60.22 && 12506 & 0 & 262.69 \\
  000550 && 346710 & 9331 & 9.88 && 305386 & 8760 & 11.21 && 41324 & 22 & 82.03 \\
  000581 && 93976 & 4471 & 35.79 && 77748 & 4192 & 43.12 && 16228 & 5 & 200.46 \\
  000625 && 397566 & 9438 & 8.43 && 350333 & 8645 & 9.56 && 47233 & 36 & 70.44 \\
  000709 && 207816 & 3676 & 16.39 && 187431 & 3461 & 18.16 && 20385 & 6 & 163.59 \\
  000720 && 132243 & 17195 & 25.42 && 110699 & 14455 & 30.35 && 21544 & 1 & 152.39 \\
  000778 && 157322 & 1527 & 21.6 && 133944 & 1318 & 25.33 && 23378 & 5 & 143.16 \\
  \hline\hline
\end{tabular}
\end{center}
\end{table}

\section{Unconditional distributions of intertrade durations}
\label{S1:UncondPDF}

\subsection{Scaling pattern}

\begin{figure}[htb]
\centering
\includegraphics[width=4.5cm]{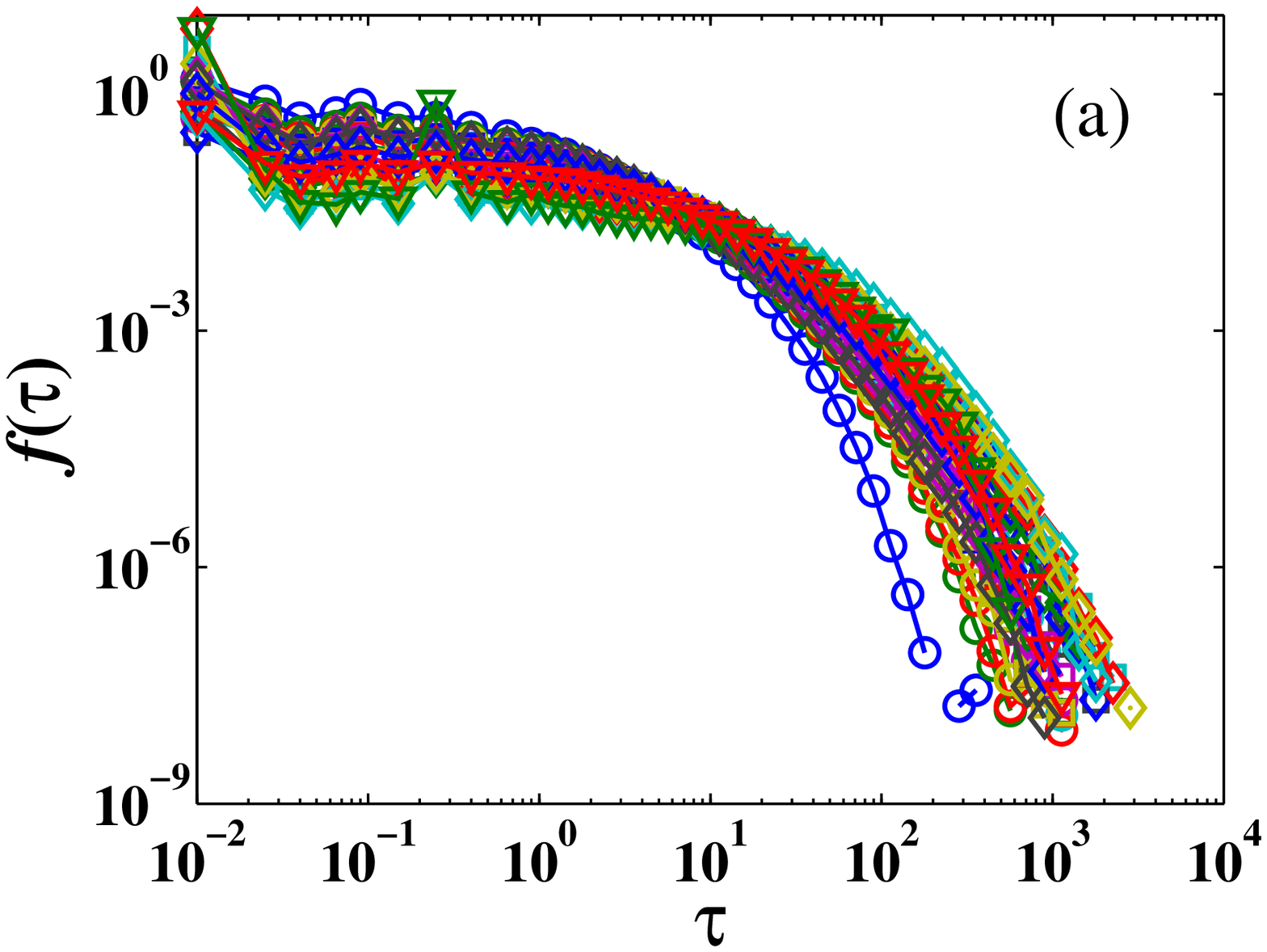}
\includegraphics[width=4.5cm]{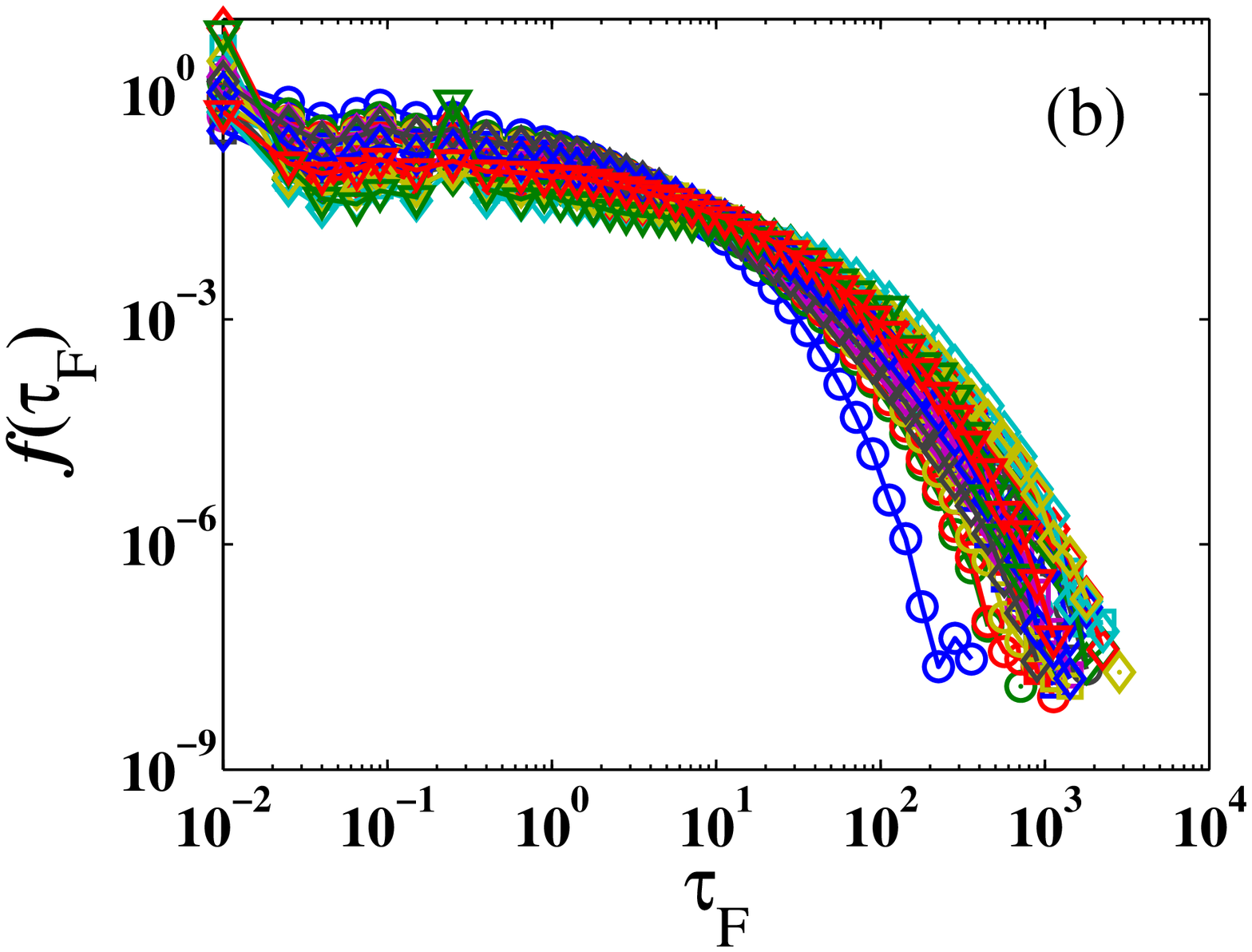}
\includegraphics[width=4.5cm]{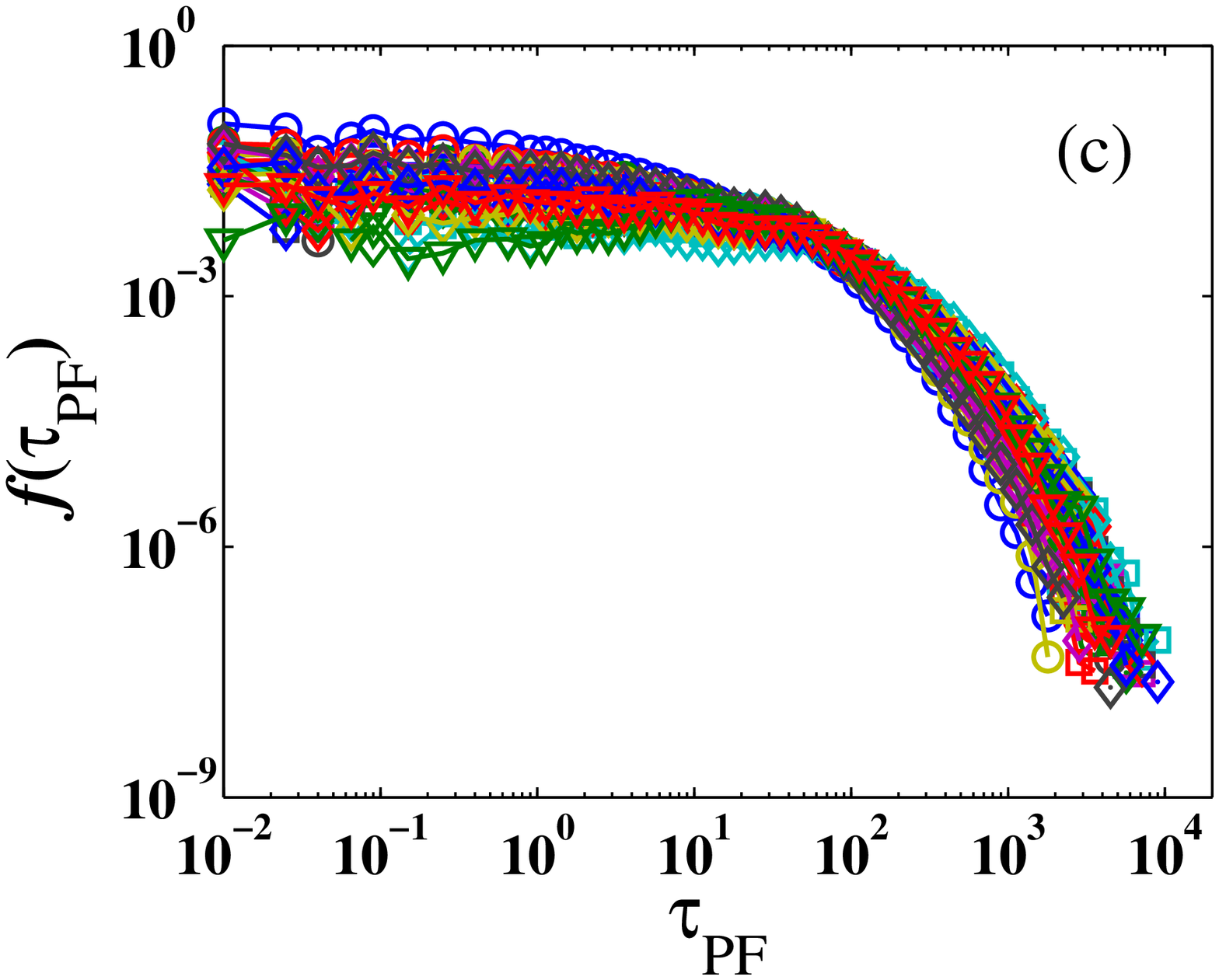}
\includegraphics[width=4.5cm]{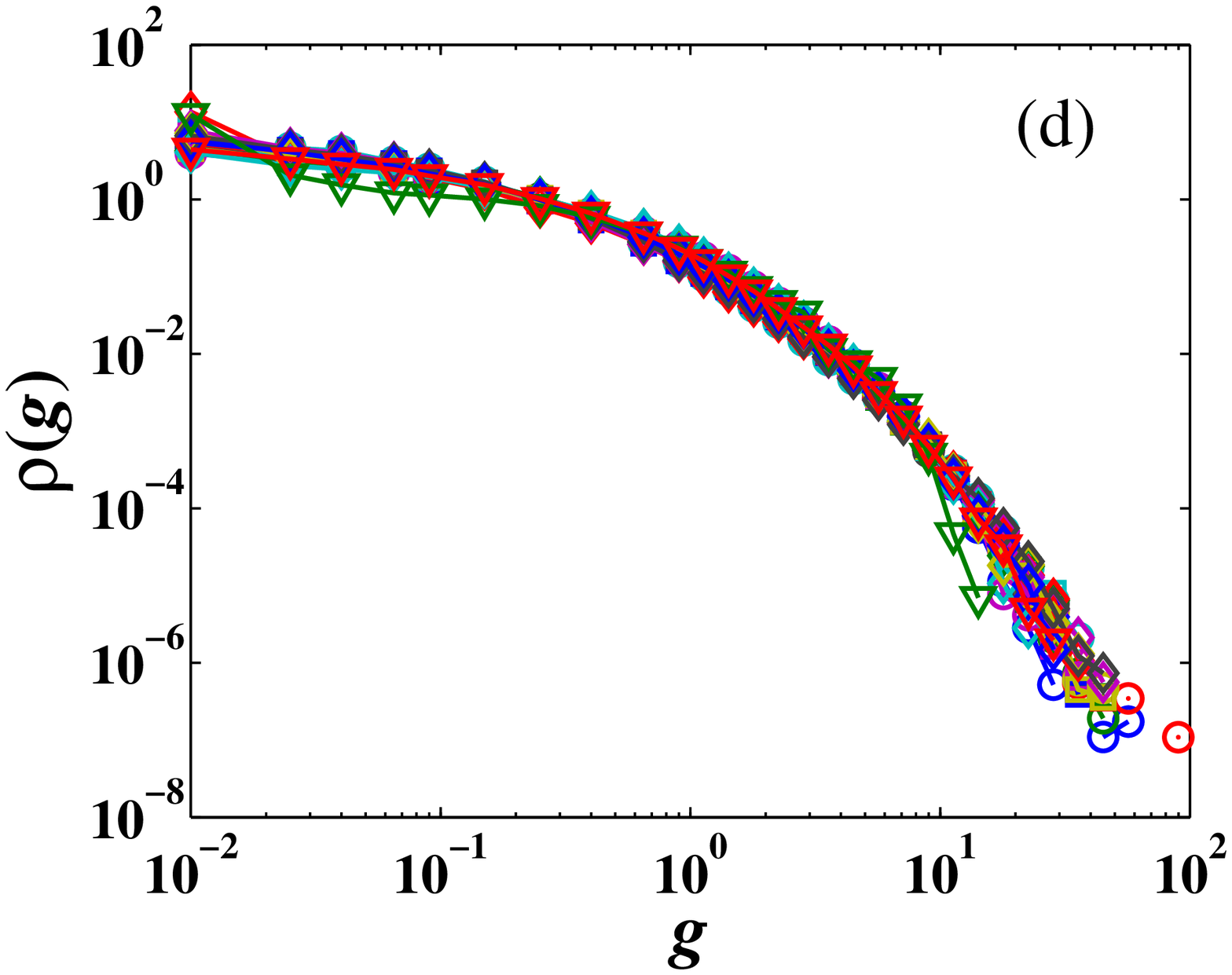}
\includegraphics[width=4.5cm]{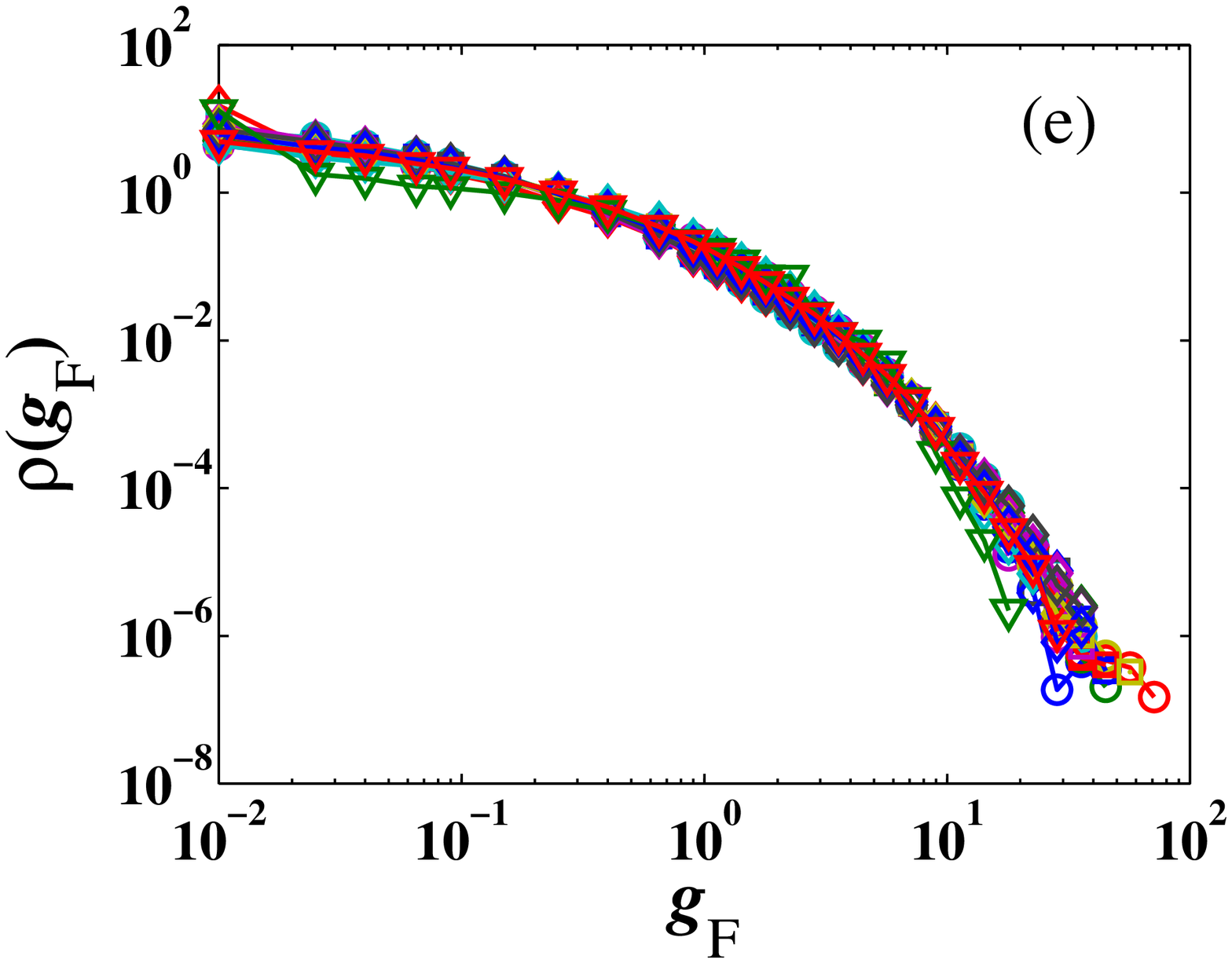}
\includegraphics[width=4.5cm]{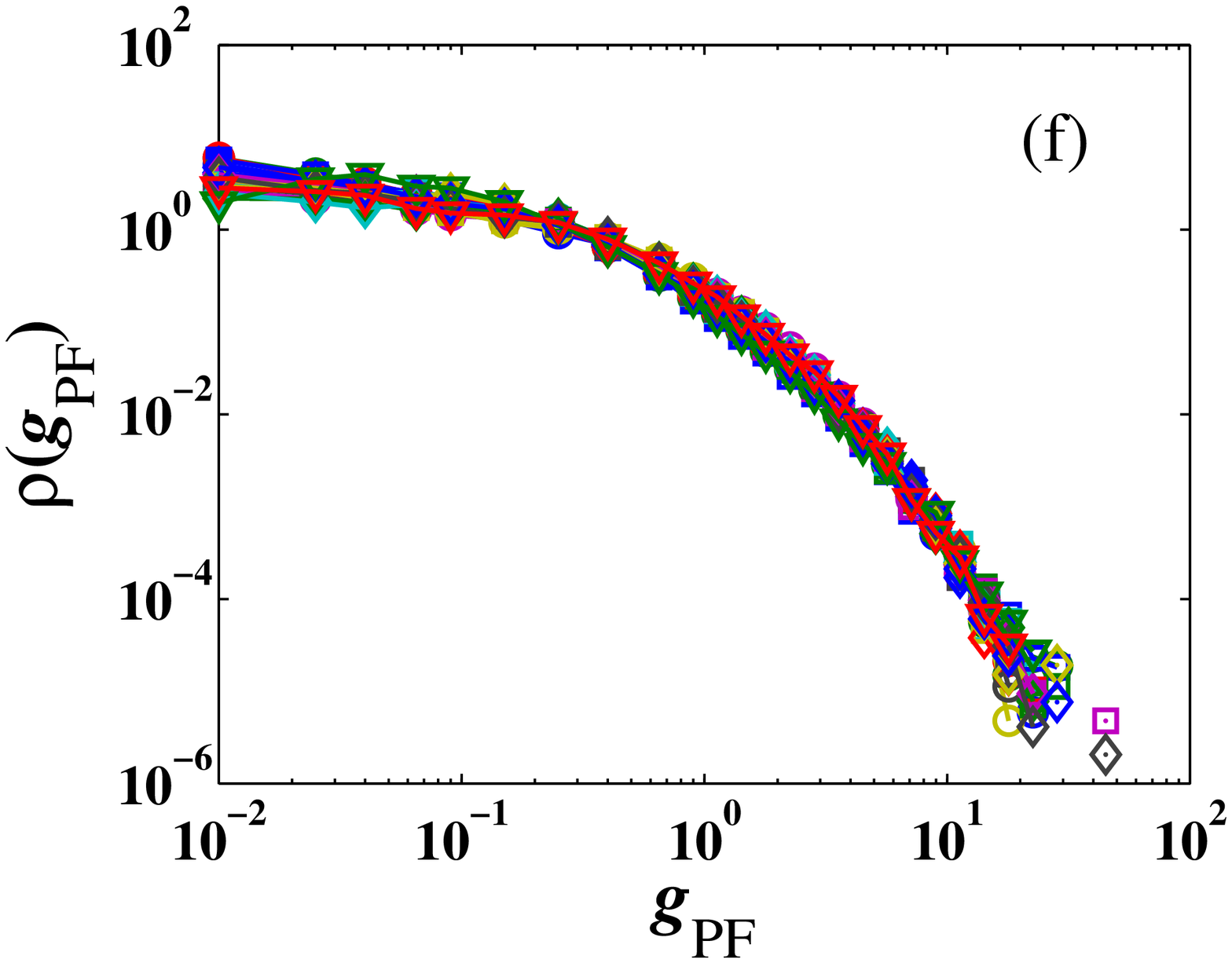}
\caption{\label{Fig:UncondPDF:Scaling} (Color online.) Scaling in
the unconditional distributions of 23 stocks. Panels (a-c)
illustrate the empirical density functions of intertrade durations
for all trades, filled trades and partially filled trades. Panels
(d-f) present the empirical density functions of the normalized
intertrade durations for all trades, filled trades and partially
filled trades.}
\end{figure}

For each stock, the three groups of durations $\tau$,
$\tau_{\rm{F}}$ and $\tau_{\rm{PF}}$ (in units of second) are
calculated for all trades, filled trades and partially filled
trades. The associated empirical density functions $f(\tau)$,
$f(\tau_{\rm{F}})$ and $f(\tau_{\rm{PF}})$ for the 23 Chinese stocks
are illustrated in the upper panel of
Fig.~\ref{Fig:UncondPDF:Scaling}. At a first glance, we observe that
different distributions share similar shapes. When the duration is
large, the tails are heavy but it is not unambiguous that they
exhibit power-law tails. Since the time stamps of the data have a
resolution of 0.01 second, the minimal duration is set to be 0.01 in
the plots. There are also vanishing durations as shown in Table
\ref{Tb:DataDetail}, which are not shown in this figure. Note that
the density functions are monotonically decreasing such that small
durations occur more frequent than large ones.

In order to compare the distributions of different stocks, we define
a normalized duration by
\begin{equation}
g = \tau/{\sigma}, ~g_{\rm{F}} = \tau_{\rm{F}}/{\sigma_{\rm{F}}},
~g_{\rm{PF}} = \tau_{\rm{PF}}/{\sigma}_{\rm{PF}},
 \label{Eq:nduration}
\end{equation}
where $\sigma$, $\sigma_{\rm{F}}$ and $\sigma_{\rm{PF}}$ are the
sample standard deviations of $\tau$, $\tau_{\rm{F}}$ and
$\tau_{\rm{PF}}$, respectively. As will be clear, the variances
$\sigma^2$, $\sigma_{\rm{F}}^2$ and $\sigma_{\rm{PF}}^2$ of
intertrade durations exist for the Chinese stocks. Then the PDFs of
these normalized durations can be obtained as follows,
\begin{equation}
\rho(g) = \sigma f(g\sigma), ~\rho(g_{\rm{F}}) = \sigma_{\rm{F}}
f(g\sigma_{\rm{F}}), ~\rho(g_{\rm{PF}}) = \sigma_{\rm{PF}}
f(g\sigma_{\rm{PF}}).
 \label{Eq:pdfnduration}
\end{equation}
In the lower panel of Fig.~\ref{Fig:UncondPDF:Scaling}, we plot
respectively $\sigma f(\tau)$, $\sigma_{\rm{F}} f(\tau_{\rm{F}})$
and $\sigma_{\rm{PF}} f(\tau_{\rm{PF}})$ as functions of
$\tau/{\sigma}$, $\tau_{\rm{F}}/{\sigma_{\rm{F}}}$ and
$\tau_{\rm{PF}}/{\sigma}_{\rm{PF}}$ for the 23 stocks. In each plot,
the 23 curves excellently collapse onto a single curve. We notice
that the collapsing for the Chinese stocks is better than those DJIA
stocks, especially for large values of the normalized durations
\cite{Ivanov-Yuen-Podobnik-Lee-2004-PRE}. This analysis shows that
$\rho(g)$, $\rho(g_{\rm{F}})$ and $\rho(g_{\rm{PF}})$ are scaling
functions.

\subsection{Fitting the intertrade duration distributions}

\begin{figure}[htb]
\centering
\includegraphics[width=4.5cm]{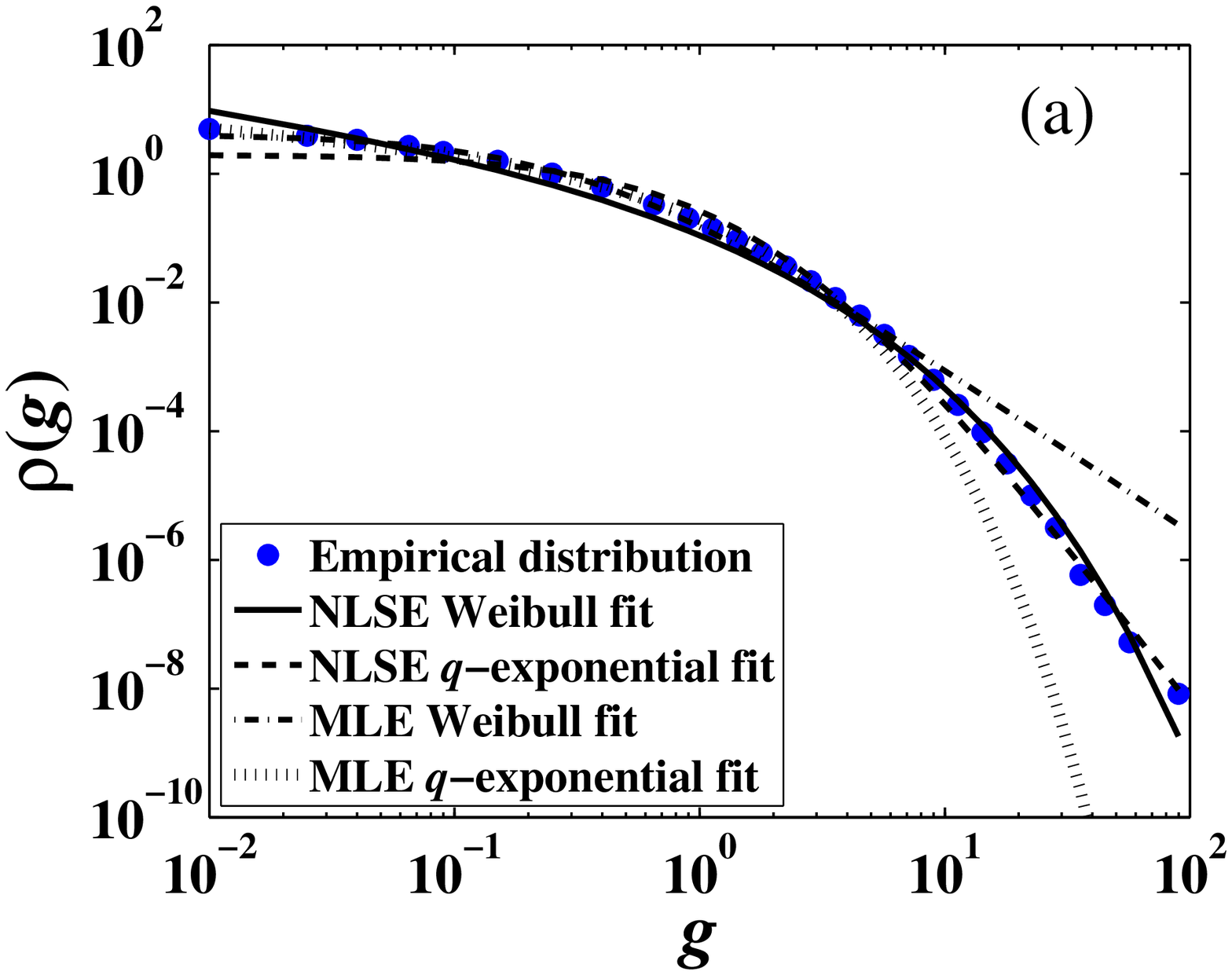}
\includegraphics[width=4.5cm]{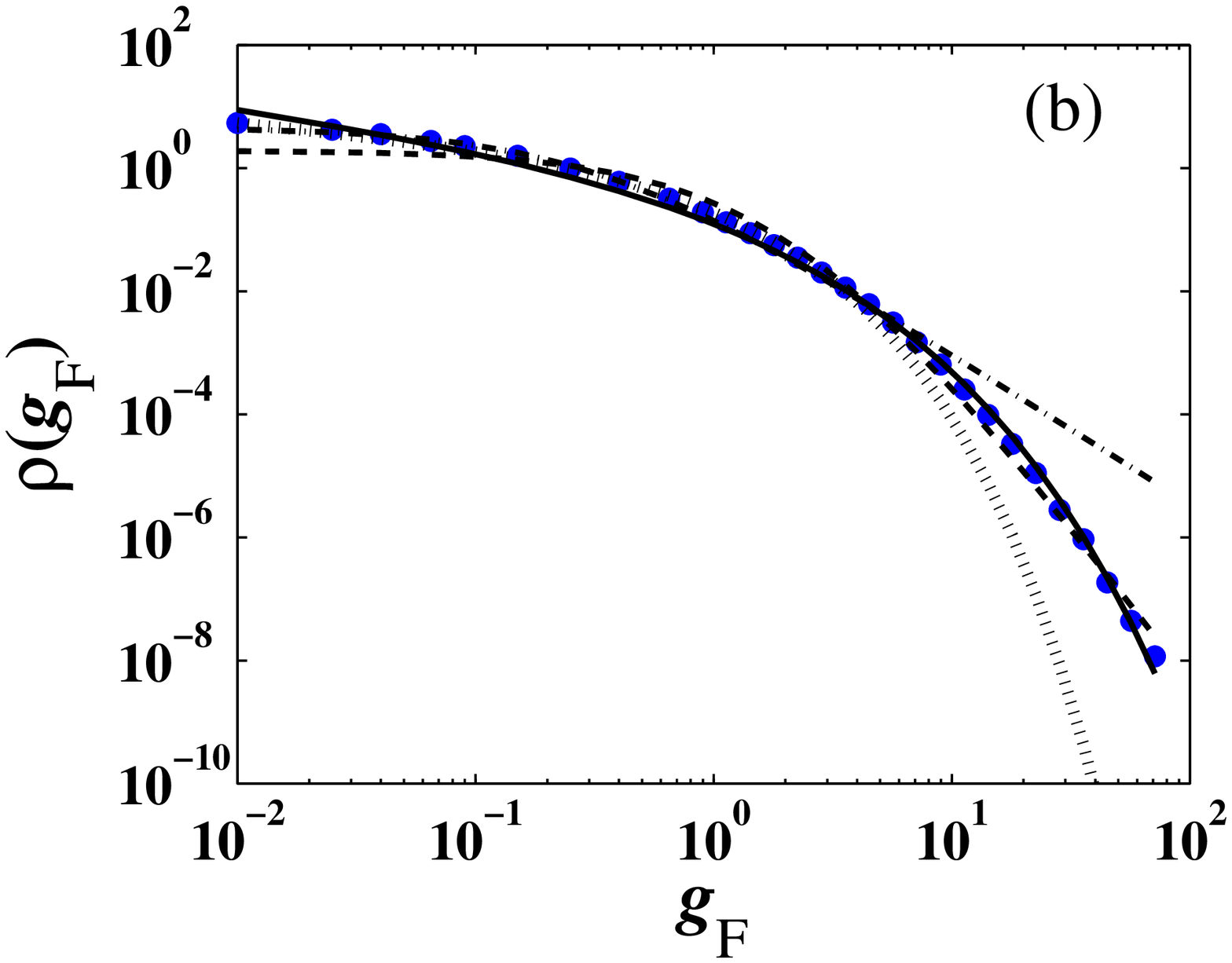}
\includegraphics[width=4.5cm]{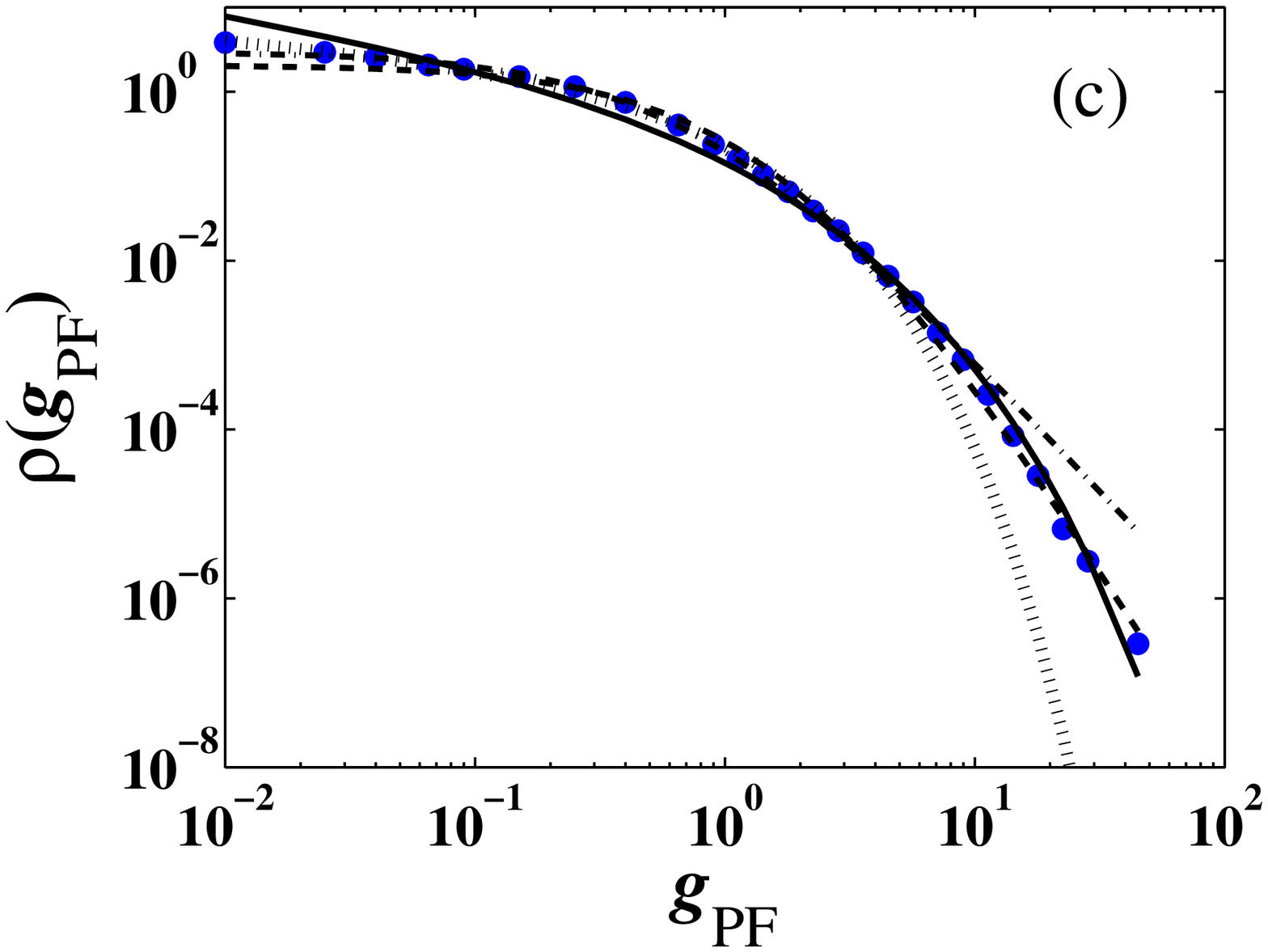}
\includegraphics[width=4.5cm]{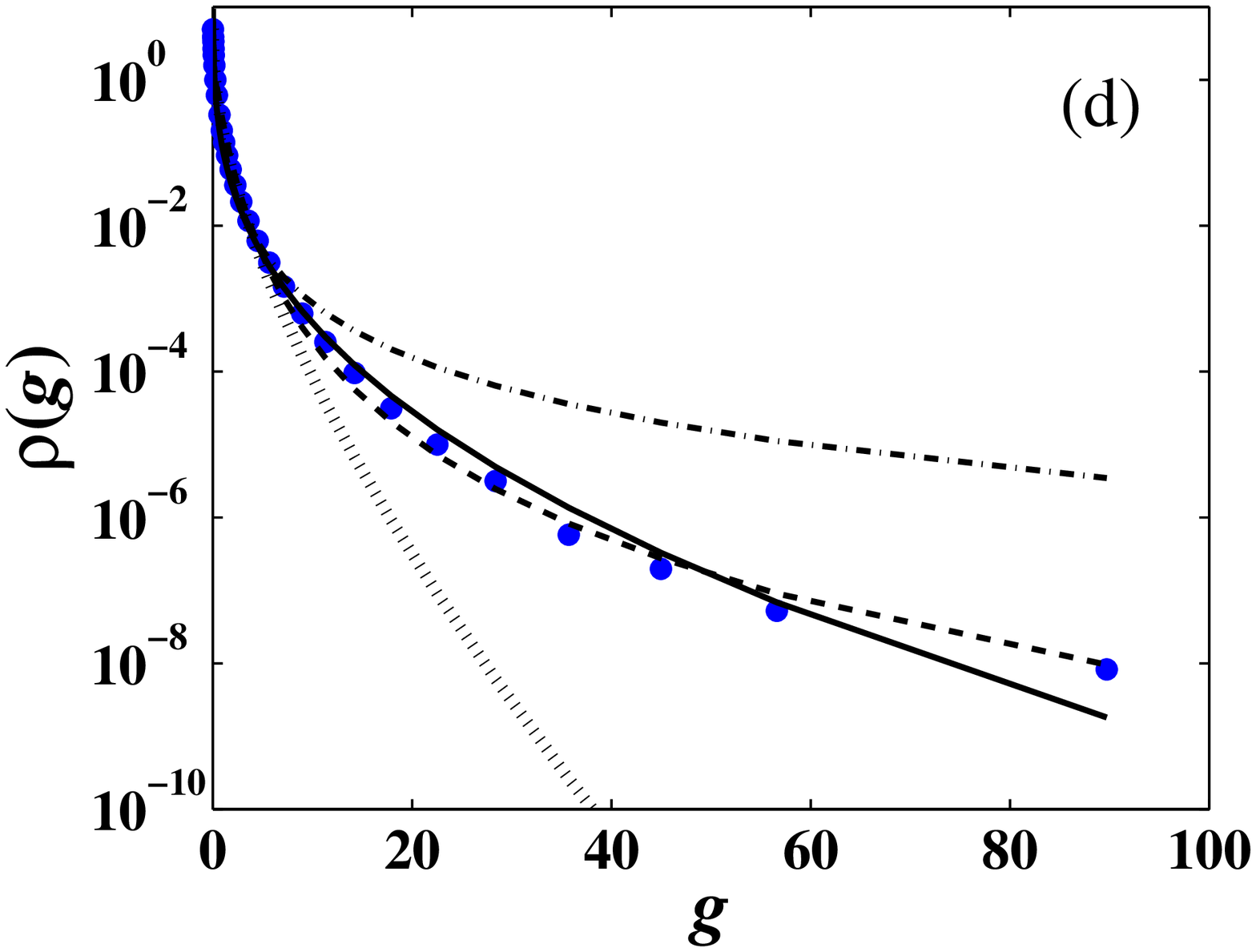}
\includegraphics[width=4.5cm]{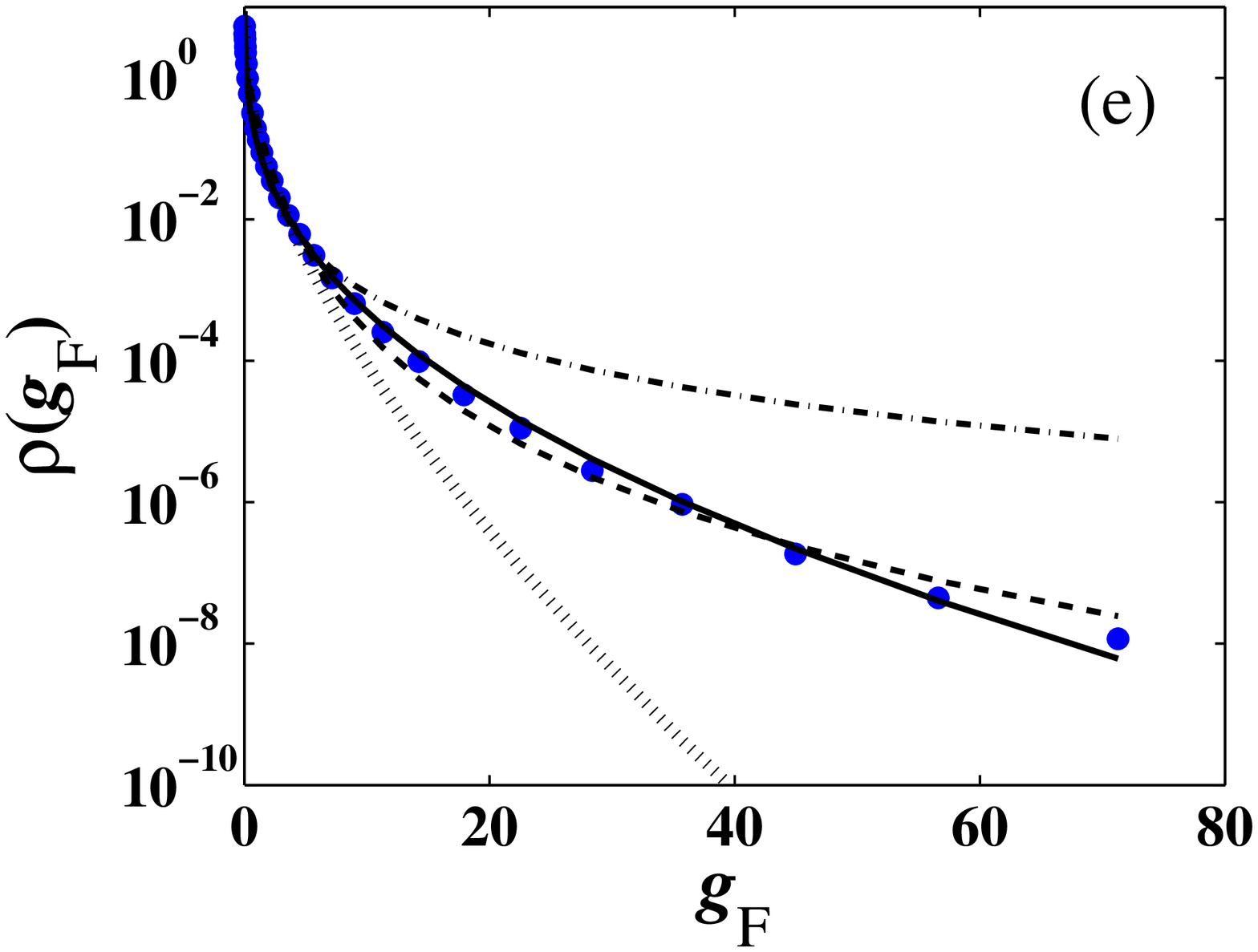}
\includegraphics[width=4.5cm]{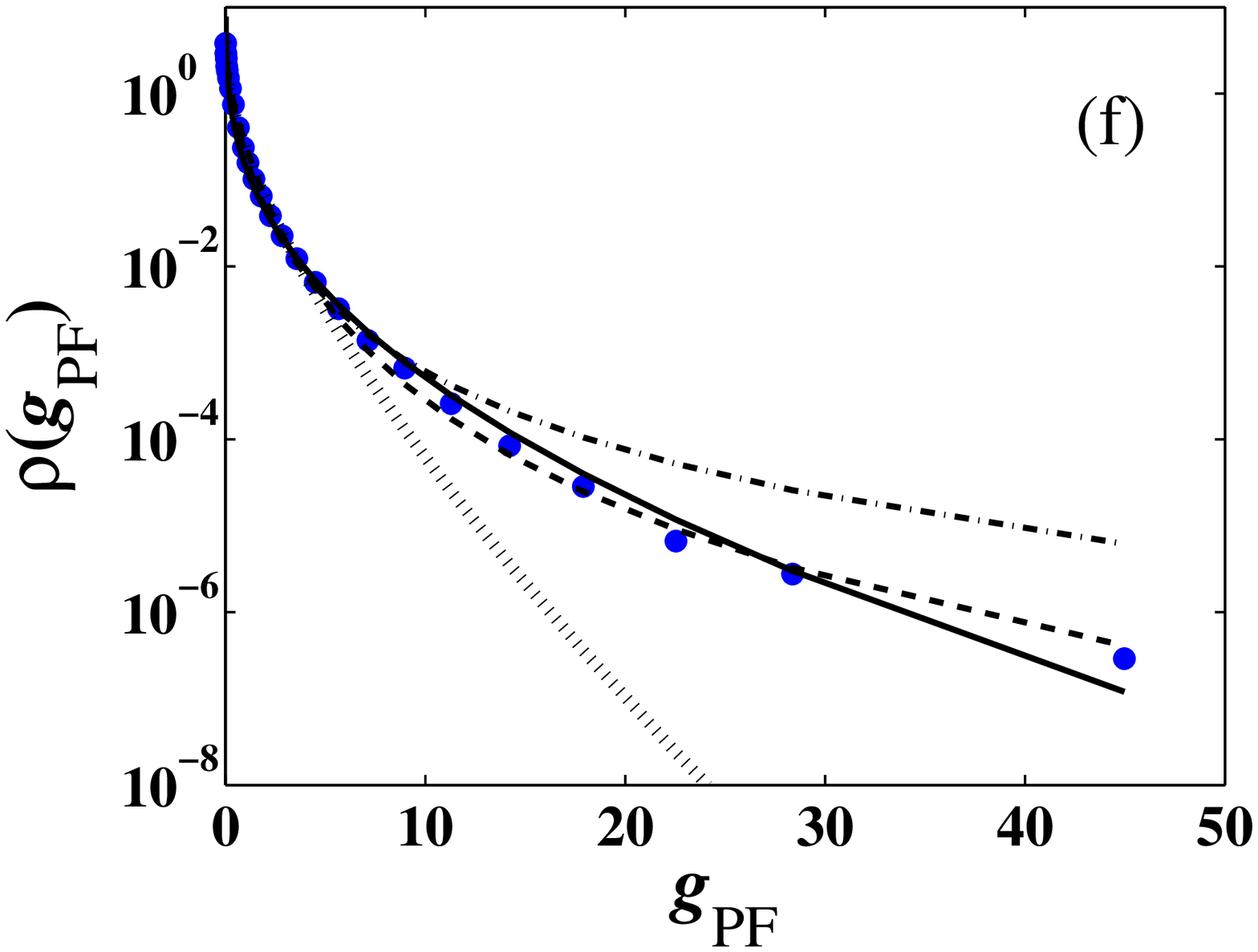}
\caption{\label{Fig:UncondPDF:Fits} (Color online.) Empirical
probability density function of normalized intertrade durations as
an ensemble of the 23 stocks. The solid and dashed lines are the
least square nonlinear fits of the Weibull and $q$-exponential
distributions. The dotted and dash-dotted lines are the maximum
likelihood fits to the Weibull and $q$-exponential distributions.
The upper panel shows double logarithmic plots, while the lower
panel shows semi-logarithmic plots for comparison.}
\end{figure}

The nice scaling in the distribution of intertrade durations means
that the normalized durations of different stocks follow the same
distribution. This allows us to treat all the normalized intertrade
durations from different stocks as an ensemble to gain better
statistics. In this way, we have three aggregate samples for $g$,
$g_{\rm{F}}$ and $g_{\rm{PF}}$. The three empirical probability
density functions of the three types of normalized intertrade
durations are illustrated in Fig.~\ref{Fig:UncondPDF:Fits} as dots.

Following the work of Politi and Scalas
\cite{Poloti-Scalas-2008-PA}, we adopt the Weibull and the
$q$-exponential distributions to model the normalized intertrade
durations. The Weibull probability density $\rho_w(g)$ can be
written as
\begin{equation}
\rho_w(g)  = \alpha \beta g^{\beta-1} \exp(-\alpha g^{\beta}),
 \label{Eq:WeibellDensity}
\end{equation}
and its complementary (cumulative) distribution function $C_w(g)$ is
\begin{equation}
C_w(g) = \exp(-\alpha g^{\beta}).
 \label{Eq:WeibellSurvival}
\end{equation}
When $\beta=1$, $\rho_w(g)$ recovers the exponential distribution.
When $0<\beta<1$, $\rho_w(g)$ is a stretched exponential or
sub-exponential. When $\beta>1$, $\rho_w(g)$ is a super-exponential.
The $q$-exponential probability density $\rho_q(g)$ is defined by
\begin{equation}
\rho_q(g)  = \mu \left[1+(1-q)(-\mu g)\right]^{\frac{q}{1-q}},
 \label{Eq:QDensity}
\end{equation}
and its complementary cumulative distribution function $C_q(\tau)$
is
\begin{equation}
C_q(g) = \left[1+(1-q)(-\mu g)\right]^{\frac{1}{1-q}}.
 \label{Eq:QSurvival}
\end{equation}
Usually, we have $q>1$. When $(1-q)(-\mu g)\gg1$, we observe a
power-law behavior in the tail $C_q(g)\sim g^{-{1}/{(q-1)}}$ with a
tail exponent of $1/(q-1)$.

In order to capture the main part of the distributions, we adopt the
maximum likelihood estimator (MLE) for model calibration.
Specifically, the MATLAB function ``wblfit'' is utilized to estimate
the parameters of the Weibull and the estimation method for the
$q$-exponential can be found in Ref.~\cite{Shalizi-2007-XXX}. The
resultant fits are illustrated in Fig.~\ref{Fig:UncondPDF:Fits}. It
is evident that both distributions fit the data very well for the
normalized durations less than about 4, which accounts for 98.5\% of
the sample. However, there are remarkable discrepancies in the tails
between the empirical data and the two models. Table~\ref{Tb:MLE}
reports the estimates of parameters. Since both Weibull and
$q$-exponential models have two parameters, we can compare
quantitatively their performance simply using the r.m.s. values of
fit residuals. According to Table \ref{Tb:MLE}, the Weibull model
outperforms the $q$-exponential model since $\chi_w<\chi_q$.

\begin{table}[htp]
 \centering
 \caption{\label{Tb:MLE} Estimated values of parameters ($\alpha$, $\beta$, $q$, $\mu$) by means of MLE.
 $\chi_w$ and $\chi_q$ stand for the r.m.s. of fit residuals.
 The $p$ value in parentheses represents the percentage of stocks preferring the chosen model in the column.}
 \medskip
 \centering
 \begin{tabular}{llllllllll}
 \hline \hline
  \multirow{3}*[3.2mm]{Duration}&& \multicolumn{3}{c}{Weibull}&&\multicolumn{3}{c}{$q$-exponential}\\  %
  \cline{3-5}  \cline{7-9}
                   &&  $\alpha$     & $\beta$    &$\chi_w$, ($p$) && $\mu$         & $q$           & $\chi_q$, ($p$) \\
    \hline
    $g$            && 1.85          & 0.68          & 0.71 && 4.17          & 1.65          & 1.35 \\
    mean $\pm$ std && $1.85\pm0.15$ & $0.69\pm0.03$ & (19/23)  && $4.18\pm0.99$ & $1.63\pm0.09$ & (4/23)  \\
    $g_{\rm{F}}$          && 1.90          & 0.67          & 0.93 && 4.60          & 1.69          & 1.57 \\
    mean $\pm$ std && $1.90\pm0.16$ & $0.67\pm0.03$ & (19/23)  && $4.64\pm1.19$ & $1.68\pm0.10$ & (4/23)  \\
    $g_{\rm{PF}}$       && 1.71          & 0.74          & 0.10 && 2.96          & 1.46          & 1.09 \\
    mean $\pm$ std && $1.73\pm0.15$ & $0.74\pm0.04$ & (14/23)  && $3.12\pm0.74$ & $1.47\pm0.10$ & (9/23)  \\
    \hline \hline
 \end{tabular}
\end{table}

We have also fitted the distributions for individual stocks (see the
three tables in Appendix of the paper at
http://arXiv.org/abs/0804.3431). For individual stocks, the Weibull
model also outperforms the $q$-exponential model. The parameters,
especially $\beta$ and $q$, are consistent across different stocks.
The mean and standard deviation for each case are also calculated
and listed in Table \ref{Tb:MLE}. We find that, for each type of
intertrade durations, the corresponding parameters estimated from
the ensemble and individual stocks are close to one another. For the
case of $g$ (and $g_{\rm{F}}$) as well, there are 19 stocks (out of
23) prefer the Weibull model in the sense that $\chi_w<\chi_q$. For
$g_{\rm{PF}}$, 14 stocks prefer the Weibull distribution and 9
stocks prefer the $q$-exponential model. Therefore, the
$q$-exponential is still a good model for the bulk distribution of
durations.

In order to capture the tail behavior of the distributions, we adopt
the nonlinear least square estimator (NLSE) for model calibration.
Specifically, the MATLAB function ``lsqcurvefit'' is utilized to
estimate the parameters of the two models. The objective function in
the fitting is $\sum[\ln\rho(g)-\ln\hat\rho(g)]^2$ rather than
$\sum[\rho(g)-\hat\rho(g)]^2$, where $\hat\rho(g)$ is the empirical
data. The resultant fits are also illustrated in
Fig.~\ref{Fig:UncondPDF:Fits} and Table~\ref{Tb:NLSE} reports the
estimates of parameters. It is evident from
Fig.~\ref{Fig:UncondPDF:Fits} that the $q$-exponential is a better
model than the Weibull except $g_{\rm{F}}$, which is confirmed by
the much smaller values of $\chi_q$ compared with $\chi_w$ in Table
\ref{Tb:NLSE}. The fact that the tail distributions can be better
modeled by the $q$-exponential implies that the intertrade durations
have a power-law tail distribution\footnote{It is easy to confirm
that the condition $(1-q)(-\mu g)\gg1$ holds when the normalized
duration is much larger than 1.}. The estimated tail exponents are
4.00, 4.17 and 3.70 for $g$, $g_{\rm{F}}$ and $g_{\rm{PF}}$.

\begin{table}[htp]
 \caption{\label{Tb:NLSE} Estimated values of parameters ($\alpha$, $\beta$, $q$, $\mu$) by means of NLSE.
  $\chi_w$ and $\chi_q$ stand for the r.m.s. of fitting residuals.
 The $p$ value in parentheses represents the percentage of stocks preferring the chosen model in the column.}
 \medskip
 \centering
 \begin{tabular}{llllllllll}
 \hline \hline
  \multirow{3}*[3.2mm]{Duration}&& \multicolumn{3}{c}{Weibull}&&\multicolumn{3}{c}{$q$-exponential}\\  %
  \cline{3-5}  \cline{7-9}
                   &&  $\alpha$     & $\beta$       & $\chi_w$, ($p$) && $\mu$& $q$ & $\chi_q$,( $p$) \\
    \hline
    $g$            && 2.24          & 0.46          & 22.41 && 1.99          & 1.25          & 17.02 \\
    mean $\pm$ std && $2.10\pm0.14$ & $0.49\pm0.03$ & (9/23)   && $2.38\pm0.29$ & $1.30\pm0.03$ & (14/23)  \\
    $g_{\rm{F}}$          && 2.12          & 0.48          & 12.44 && 1.93          & 1.24          & 23.03 \\
    mean $\pm$ std && $2.18\pm0.14$ & $0.47\pm0.04$ & (11/23)   && $2.51\pm0.38$ & $1.31\pm0.04$ & (12/23)  \\
    $g_{\rm{PF}}$       && 1.96          & 0.52          & 19.26 && 2.07          & 1.27          & 4.81 \\
    mean $\pm$ std && $2.00\pm0.14$ & $0.50\pm0.04$ & (2/23)   && $2.64\pm0.41$ & $1.36\pm0.04$ & (21/23)  \\
    \hline \hline
 \end{tabular}
\end{table}

Similarly, we have also fitted the distributions for individual
stocks using NLSE (see the three tables in Appendix of the paper at
http://arXiv.org/abs/0804.3431). For more than a half of the stocks,
the $q$-exponential model also outperforms the Weibull model. The
parameters, especially $\beta$ and $q$, are consistent across
different stocks. The mean and standard deviation for each case are
also calculated and listed in Table \ref{Tb:NLSE}. We find that, for
each type of intertrade durations, the corresponding parameters
estimated from the ensemble and individual stocks are close to one
another. The tail exponents for $g$, $g_{\rm{F}}$ and $g_{\rm{PF}}$
are estimated to be 3.33, 3.23, and 2.78, which is reminiscent of
the inverse cubic law of returns at microscopic timescale
\cite{Gopikrishnan-Meyer-Amaral-Stanley-1998-EPJB,Gu-Chen-Zhou-2008a-PA}.

\section{Conditional distributions of intertrade durations}
\label{S1:CondPDF}

\begin{figure}[htb]
\centering
\includegraphics[width=4.5cm]{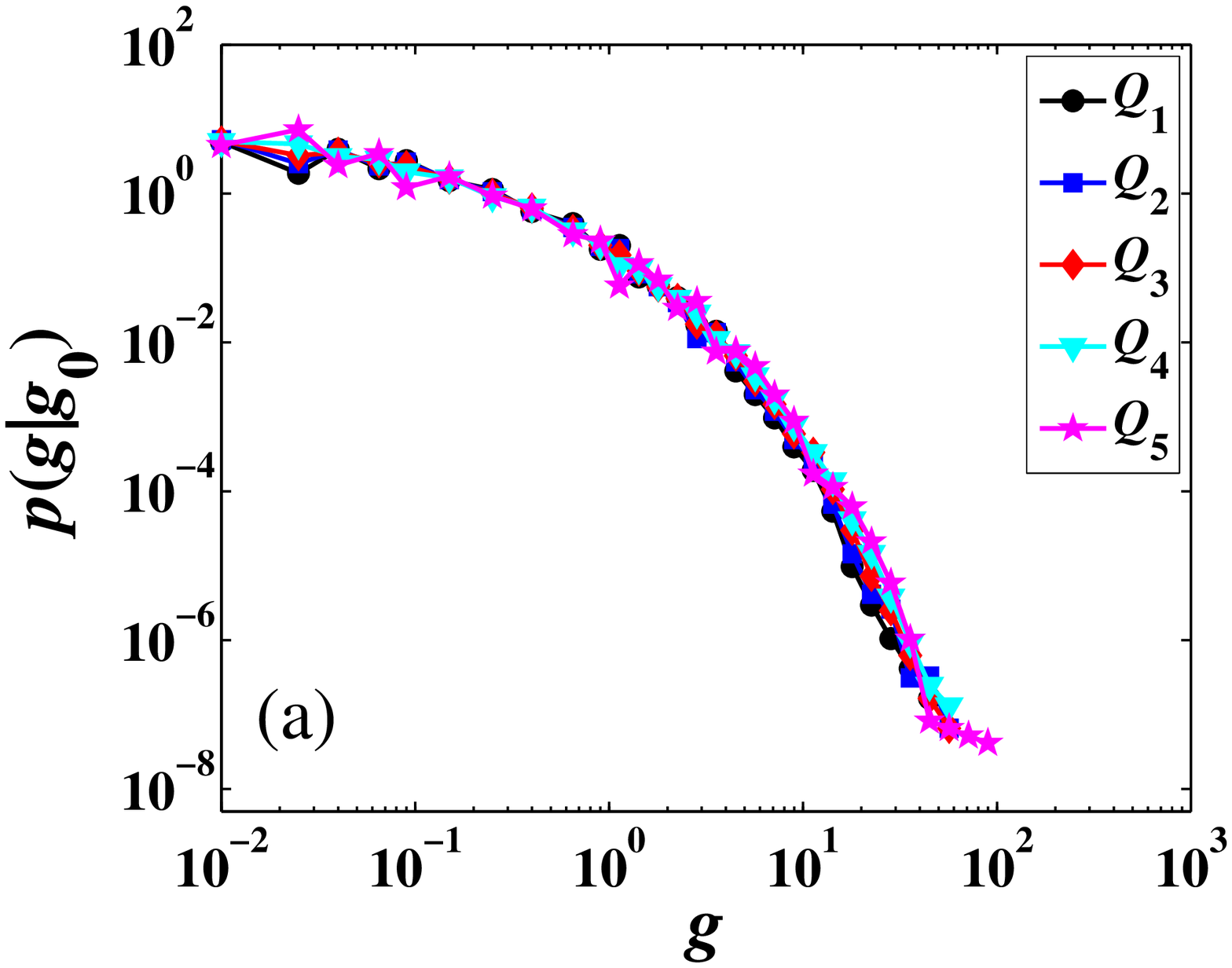}
\includegraphics[width=4.5cm]{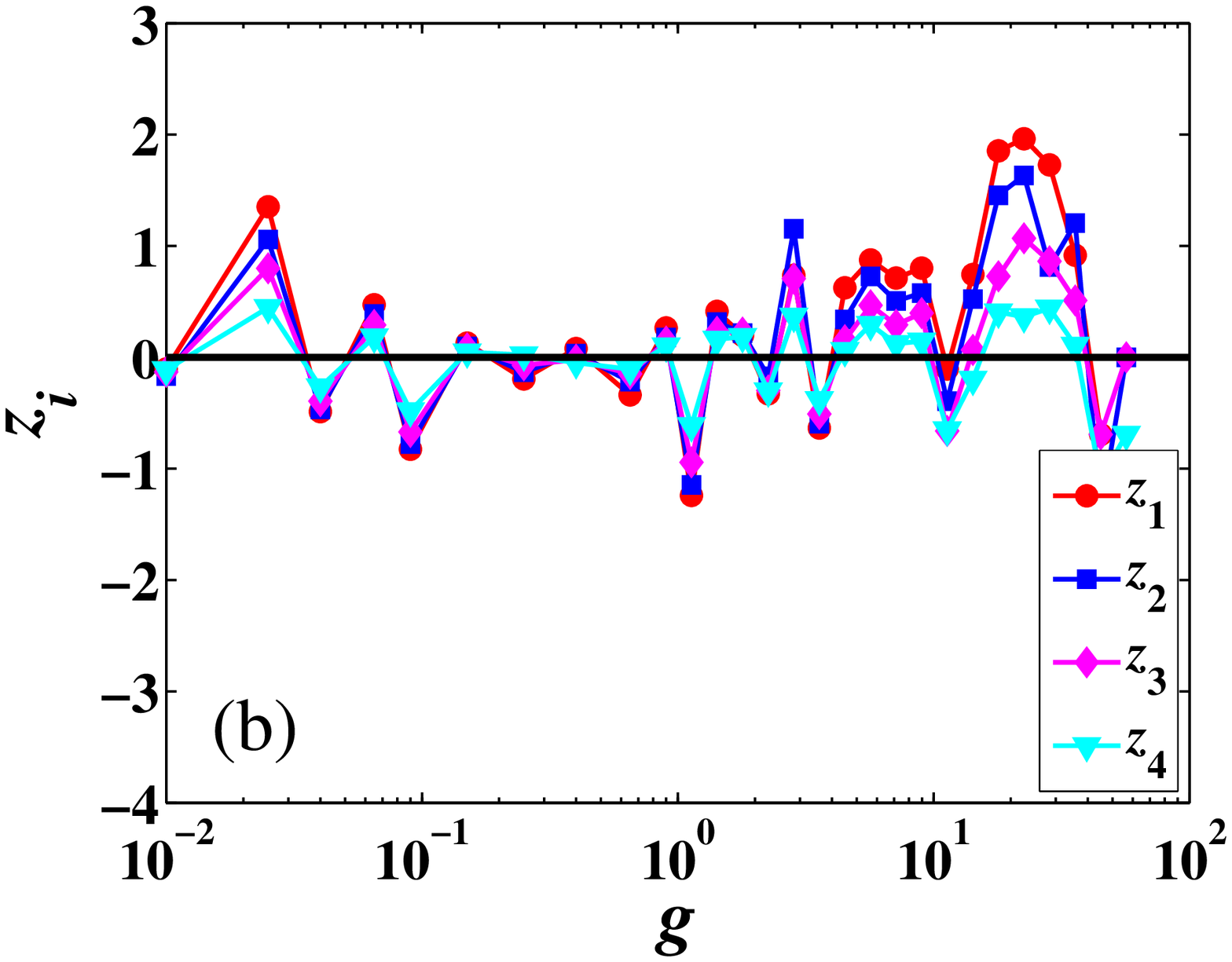}
\includegraphics[width=4.5cm]{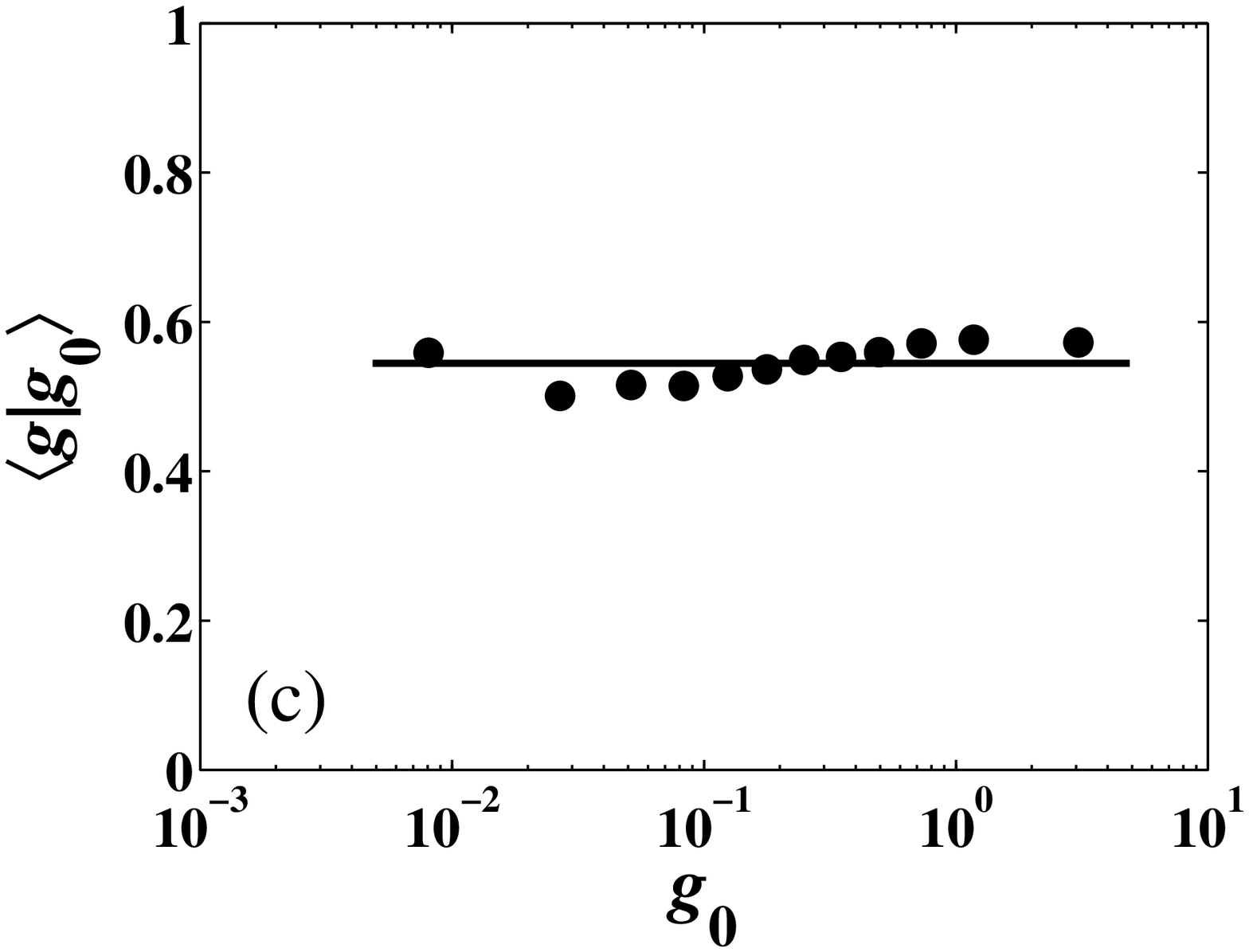}
\caption{\label{Fig:CondPDF} (Color online.) Scaling in the
conditional distributions of intertrade durations. (a) Conditional
probability distribution $p(g|g_0)$ with respect to $g$. (b) Plots
of $z_i = \ln[p(g|Q_5)/p(g|Q_i)]$ for $i=1$ to 4. (c) Mean
conditional duration $\langle g|g_0 \rangle$ with respect to $g_0$.}
\end{figure}

We now investigate the conditional distribution of normalized
intertrade durations on the value of its preceding duration. All the
normalized durations for different stocks constitute an ensemble set
$Q$, which is partitioned into five non-overlapping groups:
\begin{equation}
 Q = Q_1\cup Q_2 \cup Q_3 \cup Q_4 \cup Q_5~,
 \label{Eq:Q}
\end{equation}
where $Q_i\cap Q_j=\phi$ for $i\neq j$. In the partitioning
procedure, all values in $Q$ are sorted in increasing order and
assigned into $Q_1$ to $Q_5$ such that their sizes are approximately
identical. We estimate the empirical conditional probability density
functions $p(g|g_0)$ of normalized intertrade durations that
immediately follow a normalized intertrade duration $g_0$ belonging
to $Q_i$. The five empirical conditional PDFs are depicted in
Fig.~\ref{Fig:CondPDF}(a). Again, all the five PDFs collapse onto a
single curve, showing a nice scaling relation in the conditional
distributions of intertrade durations. Comparing with
Fig.~\ref{Fig:UncondPDF:Fits}, we see that $p(g|g_0)\approx\rho(g)$.

However, a careful scrutiny of Fig.~\ref{Fig:CondPDF}(a) unveils a
systematic trend in the five conditional PDFs identifying that, in
the tails, the PDF for $Q_1$ is on the bottom while that for $Q_5$
lies on the top. It means that there are more large durations for
$Q_i$ than $Q_j$ if $i>j$. Speaking differently, large durations
tend to follow large durations. To further clarify this observation,
we plot $z_i = \ln[p(g|Q_5)/p(g|Q_i)]$ for $i=1$ to 4 in
Fig.~\ref{Fig:CondPDF}(b). It is clear that the five PDFs deviate
from one another systematically for large $g$. The discrepancy is
much weaker for small $g$. This phenomenon can also be confirmed by
the mean conditional duration $\langle g|g_0 \rangle$. If
$p(g|g_0)=\rho(g)$, we will have
\begin{equation}
 \langle g|g_0 \rangle = \int_0^\infty p(g|g_0) g {\rm{d}}g
 = \int_0^\infty \rho(g) g  {\rm{d}}g =  \langle g \rangle~.
\end{equation}
In other words, the mean conditional duration $\langle g|g_0
\rangle$ is independent of $g_0$. Fig.~\ref{Fig:CondPDF}(c) plots
the mean conditional duration $\langle g|g_0 \rangle$ against $g_0$.
When $g_0$ is not too small, we see an upward trend in $\langle
g|g_0 \rangle$ with respect to $g_0$. However, the $\langle g|g_0
\rangle$ values are still close to $\langle g \rangle$, indicated by
the horizontal line in Fig.~\ref{Fig:CondPDF}(c).

\section{Conclusion}
\label{S1:Conclusion}

We have investigated the intertrade durations calculated from the
limit order book data of 23 liquid Chinese stocks traded on the SZSE
in the whole year 2003. The density functions are found to be
monotonically decreasing with respect to increasing intertrade
duration. After normalized by the stock-dependent standard deviation
of intertrade durations, the 23 empirical distributions of durations
collapse onto a single curve, indicating a nice scaling pattern.
This scaling behavior is also observed in the distributions of
waiting times between consecutive filled trades or partially filled
trades. Therefore, we can treat the normalized intertrade durations
of different stocks as realizations of an ensemble. The scaling
pattern implies that there are common features in the trading
behavior of market participants, which also has important
implications for market microstructure theory.

The three ensemble distributions of normalized intertrade durations
for all trades, filled trades and partially filled trades are
modeled by the Weibull and the Tsallis $q$-exponential using maximum
likelihood estimator and nonlinear least-squares estimator. We find
that more than 98.5\% of the intertrade durations can be well
modeled by the Weibull function using MLE, except for the tails, and
the logarithmic density functions can be better fitted by the
$q$-exponential function utilizing NLSE. By and large, the
intertrade duration distribution has a Weibull form followed by a
power-law tail for large durations. We also studied the conditional
distribution of normalized intertrade durations that immediately
follow a normalized intertrade duration. A scaling pattern is also
observed in these conditional distributions decorated with a weak
but systematic trend. Accordingly, the mean conditional intertrade
duration is weakly dependent on the preceding intertrade duration.

\bigskip
{\textbf{Acknowledgments:}}

We are grateful to Z. Eisler for discussion. This work was partially
supported by the National Natural Science Foundation of China (Nos.
70501011 and 70502007), the Fok Ying Tong Education Foundation (No.
101086), the Shanghai Rising-Star Program (No. 06QA14015), and the
Program for New Century Excellent Talents in University (No.
NCET-07-0288).

\bibliography{E:/Papers/Auxiliary/Bibliography}

\begin{thebibliography}{10}
\expandafter\ifx\csname url\endcsname\relax
  \def\url#1{\texttt{#1}}\fi
\expandafter\ifx\csname urlprefix\endcsname\relax\def\urlprefix{URL }\fi

\bibitem{Diamond-Verrecchia-1987-JFE}
D.~W. Diamand, R.~E. Verrecchia, {Constraints on short-selling and asset price
  adjustment to private information}, J. Finan. Econ. 18 (1987) 277--311.

\bibitem{Easley-OHara-1992-JF}
D.~Easley, M.~O'Hara, {Time and the process of security price adjustment}, J.
  Finan. 47 (1992) 577--605.

\bibitem{Dufour-Engle-2000-JF}
A.~Dufour, R.~F. Engle, {Time and the price impact of a trade}, J. Finan. 55
  (2000) 2467--2498.

\bibitem{Jasiak-1999-Finance}
J.~Jasiak, {Persistence in Intertrade durations}, Finance 19 (1998) 166--195.

\bibitem{Engle-Russell-1998-Em}
R.~Engle, J.~R. Russell, {The Autoregressive Conditional Duration Model},
  Econometrica 66 (1998) 1127--1163.

\bibitem{Engle-2000-Em}
R.~F. Engle, {The econometrics of ultra-high-frequency data}, Econometrica 68
  (2000) 1--20.

\bibitem{Bauwens-Giot-2000-AES}
L.~Bauwens, P.~Giot, {The logarithmic ACD model: An application to the bid-ask
  quote process of three NYSE stocks}, Ann. Econ. Stat. 60 (2000) 117--149.

\bibitem{Zhang-Russell-Tsay-2001-JEm}
M.~Y.~J. Zhang, J.~R. Russell, R.~S. Tsay, {A nonlinear autoregressive
  conditional duration model with applications to financial transaction data},
  J. Econometrics 104 (2001) 179--207.

\bibitem{Bauwens-Veredas-2004-JEm}
L.~Bauwens, D.~Veredas, {The stochastic conditional duration model: A latent
  variable model for the analysis of financial durations}, J. Econometrics 119
  (2004) 381--412.

\bibitem{Ghysels-Gourieroux-Jasiak-2004-JEm}
E.~Ghysels, C.~Gourieroux, J.~Jasiak, {Stochastic volatility duration models},
  J. Econometrics 119 (2004) 413--433.

\bibitem{Sun-Rachev-Fabozzi-Kalev-2008-AF}
W.~Sun, S.~Rachev, F.~J. Fabozzi, P.~S. Kalev, {Fractals in trade duration:
  capturing long-range dependence and heavy tailedness in modeling trade
  duration}, Ann. Finan. 4 (2008) 217--241.

\bibitem{Burr-1942-AMS}
I.~W. Burr, {Cumulative frequency functions}, Ann. Math. Stat. 13 (1942)
  215--232.

\bibitem{Tsallis-1988-JSP}
C.~Tsallis, {Possible generalization of Boltzmann-Gibbs statistics}, J. Stat.
  Phys. 52 (1988) 479--487.

\bibitem{Nadarajah-Kotz-2006-PLA}
S.~Nadarajah, S.~Kotz, {$q$ exponential is a Burr distribution}, Phys. Lett. A
  359 (2006) 577--579.

\bibitem{Scalas-Gorenflo-Mainardi-2000-PA}
E.~Scalas, R.~Gorenflo, F.~Mainardi, {Fractional calculus and continuous-time
  finance}, Physica A 284 (2000) 376--384.

\bibitem{Mainardi-Raberto-Gorenflo-Scalas-2000-PA}
F.~Mainardi, M.~Raberto, R.~Gorenflo, E.~Scalas, {Fractional calculus and
  continuous-time finance II: The waiting-time distribution}, Physica A 287
  (2000) 468--481.

\bibitem{Masoliver-Montero-Weiss-2003-PRE}
J.~Masoliver, M.~Montero, G.~H. Weiss, {Continuous-time random walk model for
  financial distribution}, Phys. Rev. E 67 (2003) 021112.

\bibitem{Scalas-2006-PA}
E.~Scalas, The application of continuous-time random walks in finance and
  economics, Physica A 362 (2006) 225--239.

\bibitem{Masoliver-Montero-Perello-Weiss-2006-JEBO}
J.~Masoliver, M.~Montero, J.~Perell{\'o}, G.~H. Weiss, The continunous time
  random walk formalism in financial markets, J. Econ. Behav. Org. 61 (2006)
  577--598.

\bibitem{Sabatelli-Keating-Dudley-Richmond-2002-EPJB}
L.~Sabatelli, S.~Keating, J.~Dudley, P.~Richmond, {Waiting time distributions
  in financial markets}, Eur. Phys. J. B 27 (2002) 273--275.

\bibitem{Raberto-Scalas-Mainardi-2002-PA}
M.~Raberto, E.~Scalas, F.~Mainardi, {Waiting-times and returns in
  high-frequency financial data: An empirical study}, Physica A 314 (2002)
  749--755.

\bibitem{Yoon-Choi-Lee-Yum-Kim-2006-PA}
S.-M. Yoon, J.~S. Choi, C.~C. Lee, M.-K. Yum, K.~Kim, {Dynamical Volatilities
  for yen-dollar exchange rates}, Physica A 59 (2006) 569--575.

\bibitem{Bartiromo-2004-PRE}
R.~Bartiromo, {Dynamic of stock price}, Phys. Rev. E 60 (2004) 067108.

\bibitem{Ivanov-Yuen-Podobnik-Lee-2004-PRE}
P.~C. Ivanov, A.~Yuen, B.~Podobnik, Y.-K. Lee, {Common scaling patterns in
  intertrade times of U. S. stocks}, Phys. Rev. E 69 (2004) 056107.

\bibitem{Sazuka-2007-PA}
N.~Sazuka, {On the gap between an empirical distribution and an exponential
  distribution of waiting times for price changes in a financial market},
  Physica A 376 (2007) 500--506.

\bibitem{Kim-Yoon-2003-Fractals}
K.~Kim, S.-M. Yoon, {Dynamic behavior of continuous tick data in futures
  exchange market}, Fractals 11~(2) (2003) 131--136.

\bibitem{Kim-Yoon-Kim-Lee-Scalas-2007-JKPS}
K.~Kim, S.-M. Yoon, S.~Y. Kim, D.-I. Lee, E.~Scalas, {Dynamical Mechanisms of
  the Continuous-Time Random Walk, Multifractals,Herd Behaviors and Minority
  Games in Financial Markets}, J. Korean Phys. Soc. 50 (2007) 182--190.

\bibitem{Scalas-Gorenflo-Luckock-Mainardi-Mantelli-Raberto-2004-QF}
E.~Scalas, R.~Gorenflo, H.~Luckock, F.~Mainardi, M.~Mantelli, M.~Raberto,
  {Anomalous waiting times in high-frequency financial data}, Quant. Finance 4
  (2004) 695--702.

\bibitem{Scalas-Gorenflo-Luckock-Mainardi-Mantelli-Raberto-2005-FL}
E.~Scalas, R.~Gorenflo, H.~Luckock, F.~Mainardi, M.~Mantelli, M.~Raberto, {On
  the Intertrade Waiting-time Distribution}, Finan. Lett. 3 (2005) 695--702.

\bibitem{Poloti-Scalas-2008-PA}
M.~Politi, E.~Scalas, {Fitting the empirical distribution of intertrade
  durations}, Physica A 387 (2008) 2025--2034.

\bibitem{Scalas-Kaizoji-Kirchler-Huber-Tedeschi-2006-PA}
E.~Scalas, T.~Kaizoji, M.~Kirchler, J.~Huber, A.~Tedeschi, {Waiting times
  between orders and trades in double-auction markets}, Physica A 366 (2006)
  463--471.

\bibitem{Politi-Scalas-2007-PA}
M.~Politi, E.~Scalas, {Activity spectrum from waiting-time distribution},
  Physica A 383 (2007) 43--48.

\bibitem{Eisler-Kertesz-2006-EPJB}
Z.~Eisler, J.~Kert{\'e}sz, {Size matters: Some stylized facts of the stock
  market revisited}, Eur. Phys. J. B 51 (2006) 145--154.

\bibitem{Zhou-2007-XXX}
W.-X. Zhou, {Universal price impact functions of individual trades in an
  order-driven market}, http://arxiv.org/abs/0708.3198v2 (2007).

\bibitem{Gu-Chen-Zhou-2007-EPJB}
G.-F. Gu, W.~Chen, W.-X. Zhou, {Quantifying bid-ask spreads in the Chinese
  stock market using limit-order book data: Intraday pattern, probability
  distribution, long memory, and multifractal nature}, Eur. Phys. J. B 57
  (2007) 81--87.

\bibitem{Gu-Chen-Zhou-2008a-PA}
G.-F. Gu, W.~Chen, W.-X. Zhou, {Empirical distributions of Chinese stock
  returns at different microscopic timescales}, Physica A 387 (2008) 495--502.

\bibitem{Gu-Chen-Zhou-2008b-PA}
G.-F. Gu, W.~Chen, W.-X. Zhou, {Empirical regularities of order placement in
  the Chinese stock market}, Physica A 387 (2008) 3173--3182.

\bibitem{Shalizi-2007-XXX}
C.~R. Shalizi, {Maximum Likelihood Estimation for $q$-Exponential (Tsallis)
  Distributions}, arXiv:math/0701854v2 (2007).

\bibitem{Gopikrishnan-Meyer-Amaral-Stanley-1998-EPJB}
P.~Gopikrishnan, M.~Meyer, L.~A.~N. Amaral, H.~E. Stanley, {Inverse cubic law
  for the distribution of stock price variations}, Eur. Phys. J. B 3 (1998)
  139--140.

\end{thebibliography}


\newpage

\appendix
\section{Fitted parameters for the three classes of intertrade durations}

\begin{table}[htp]
 \centering
 \caption{\label{Tb:FitValuesAll} Values of fitted parameters,
$\alpha$, $\beta$ $q$, $\mu$ for the distribution of 23 stocks for
all trades duration.}
 \medskip
 \centering
 \begin{tabular}{lccccccccccccc}
 \hline \hline
    Stock code  &  \multicolumn{6} {@{\extracolsep\fill}c}{MLE} & & \multicolumn{6}{@{\extracolsep\fill}c}{NLSE} \\
    \cline{2-7} \cline{9-14}
    &  $\alpha$ & $\beta$ & $\chi_w$ & $\mu$ & $q$ & $\chi_q$ & & $\alpha$ & $\beta$ & $\chi_w$ & $\mu$ & $q$ & $\chi_q$ \\
    \hline
    000001 & 1.69 & 0.71 & 1.04 & 3.30 & 1.57 & 1.11 && 2.15 & 0.49 & 27.17 & 1.99 & 1.24 & 8.67 \\
    000002 & 1.89 & 0.68 & 0.89 & 4.41 & 1.67 & 1.08 && 2.01 & 0.51 & 10.11 & 2.17 & 1.27 & 16.18 \\
    000009 & 1.84 & 0.69 & 0.57 & 3.95 & 1.61 & 1.25 && 2.42 & 0.41 & 35.64 & 2.58 & 1.32 & 8.12 \\
    000012 & 2.14 & 0.66 & 2.09 & 5.72 & 1.70 & 0.57 && 2.44 & 0.40 & 25.37 & 3.12 & 1.38 & 14.96 \\
    000016 & 1.63 & 0.73 & 0.38 & 2.88 & 1.50 & 1.00 && 2.38 & 0.42 & 50.14 & 2.38 & 1.30 & 2.85 \\
    000021 & 1.88 & 0.68 & 1.14 & 4.34 & 1.66 & 1.40 && 2.07 & 0.49 & 12.97 & 2.34 & 1.29 & 14.54 \\
    000024 & 1.89 & 0.70 & 0.65 & 4.00 & 1.59 & 0.95 && 2.05 & 0.49 & 16.14 & 2.38 & 1.30 & 9.92 \\
    000027 & 2.01 & 0.65 & 1.58 & 5.48 & 1.75 & 0.95 && 2.10 & 0.49 & 9.06 & 2.46 & 1.30 & 21.85 \\
    000063 & 1.87 & 0.68 & 0.83 & 4.20 & 1.64 & 1.08 && 2.03 & 0.50 & 13.23 & 2.38 & 1.31 & 11.85 \\
    000066 & 1.87 & 0.68 & 0.77 & 4.22 & 1.65 & 1.41 && 2.04 & 0.50 & 11.92 & 2.32 & 1.29 & 13.43 \\
    000088 & 1.80 & 0.70 & 0.21 & 3.59 & 1.55 & 1.46 && 2.01 & 0.50 & 16.18 & 2.36 & 1.31 & 7.29 \\
    000089 & 1.85 & 0.67 & 0.54 & 4.23 & 1.66 & 2.57 && 1.93 & 0.53 & 4.76 & 2.21 & 1.28 & 16.79 \\
    000406 & 1.76 & 0.72 & 0.46 & 3.37 & 1.53 & 0.87 && 1.99 & 0.51 & 17.43 & 2.09 & 1.26 & 7.68 \\
    000429 & 1.91 & 0.67 & 0.84 & 4.62 & 1.68 & 1.30 && 2.06 & 0.50 & 10.13 & 2.27 & 1.28 & 17.15 \\
    000488 & 1.77 & 0.73 & 0.41 & 3.19 & 1.48 & 0.77 && 1.99 & 0.51 & 20.59 & 2.18 & 1.28 & 5.37 \\
    000539 & 2.00 & 0.62 & 2.87 & 5.98 & 1.82 & 8.68 && 2.07 & 0.48 & 1.67 & 2.55 & 1.33 & 34.82 \\
    000541 & 1.63 & 0.74 & 0.19 & 2.77 & 1.46 & 1.01 && 1.98 & 0.51 & 23.16 & 2.05 & 1.27 & 4.04 \\
    000550 & 2.07 & 0.64 & 1.76 & 6.02 & 1.78 & 1.46 && 2.18 & 0.46 & 8.91 & 2.91 & 1.36 & 22.04 \\
    000581 & 1.88 & 0.68 & 0.62 & 4.26 & 1.64 & 1.31 && 2.09 & 0.48 & 14.93 & 2.46 & 1.31 & 11.66 \\
    000625 & 2.06 & 0.67 & 1.39 & 5.25 & 1.69 & 0.74 && 2.17 & 0.46 & 14.72 & 2.9 & 1.36 & 13.94 \\
    000709 & 1.86 & 0.68 & 1.05 & 4.28 & 1.66 & 1.31 && 2.09 & 0.49 & 14.62 & 2.37 & 1.30 & 13.29 \\
    000720 & 1.53 & 0.67 & 7.55 & 2.75 & 1.55 & 19.08 && 2.04 & 0.50 & 11.30 & 2.01 & 1.26 & 24.38 \\
    000778 & 1.74 & 0.71 & 0.46 & 3.34 & 1.54 & 0.93 && 2.04 & 0.50 & 21.34 & 2.21 & 1.28 & 6.48 \\
    \hline \hline
 \end{tabular}
\end{table}

\begin{table}[htp]
 \centering
 \caption{\label{Tb:FitValuesFilled} Values of fitted parameters,
$\alpha$, $\beta$ $q$, $\mu$ for the distribution of 23 stocks for
filled trades duration.}
 \medskip
 \centering
 \begin{tabular}{lccccccccccccc}
 \hline \hline
    Stock code  &  \multicolumn{6} {@{\extracolsep\fill}c}{MLE} & & \multicolumn{6}{@{\extracolsep\fill}c}{NLSE} \\
    \cline{2-7} \cline{9-14}
    &  $\alpha$ & $\beta$ & $\chi_w$ & $\mu$ & $q$ & $\chi_q$ & & $\alpha$ & $\beta$ & $\chi_w$ & $\mu$ & $q$ & $\chi_q$ \\
    \hline
    000001 & 1.73 & 0.70 & 0.85 & 3.51 & 1.59 & 1.17 && 2.01 & 0.53 & 14.82 & 1.93 & 1.23 & 11.48 \\
    000002 & 1.94 & 0.67 & 1.02 & 4.80 & 1.69 & 1.21 && 2.03 & 0.51 & 8.65 & 2.25 & 1.28 & 18.92 \\
    000009 & 1.87 & 0.68 & 0.83 & 4.17 & 1.64 & 1.52 && 2.34 & 0.43 & 27.65 & 2.49 & 1.31 & 11.52 \\
    000012 & 2.21 & 0.64 & 3.43 & 6.52 & 1.76 & 0.81 && 2.38 & 0.42 & 18.42 & 2.96 & 1.35 & 25.53 \\
    000016 & 1.68 & 0.71 & 0.39 & 3.20 & 1.55 & 1.23 && 2.26 & 0.44 & 36.21 & 2.28 & 1.29 & 5.61 \\
    000021 & 1.95 & 0.66 & 1.44 & 4.93 & 1.71 & 1.46 && 2.30 & 0.44 & 20.13 & 2.61 & 1.32 & 16.78 \\
    000024 & 1.93 & 0.68 & 0.82 & 4.45 & 1.64 & 1.09 && 2.28 & 0.44 & 24.24 & 2.52 & 1.32 & 12.35 \\
    000027 & 2.07 & 0.64 & 1.80 & 6.04 & 1.79 & 1.03 && 2.33 & 0.43 & 15.05 & 2.69 & 1.33 & 23.85 \\
    000063 & 1.93 & 0.67 & 1.08 & 4.69 & 1.69 & 1.19 && 2.27 & 0.44 & 21.13 & 2.48 & 1.31 & 15.25 \\
    000066 & 1.93 & 0.66 & 1.18 & 4.77 & 1.70 & 1.50 && 2.31 & 0.44 & 20.41 & 2.44 & 1.30 & 17.45 \\
    000088 & 1.82 & 0.69 & 0.25 & 3.87 & 1.60 & 1.78 && 2.23 & 0.45 & 24.79 & 2.33 & 1.30 & 10.15 \\
    000089 & 1.91 & 0.65 & 1.16 & 4.85 & 1.73 & 3.27 && 2.28 & 0.44 & 14.34 & 2.45 & 1.31 & 21.15 \\
    000406 & 1.83 & 0.70 & 0.53 & 3.77 & 1.58 & 1.02 && 2.16 & 0.47 & 24.05 & 2.36 & 1.30 & 8.39 \\
    000429 & 1.96 & 0.65 & 1.12 & 5.12 & 1.73 & 1.49 && 2.24 & 0.46 & 14.71 & 2.61 & 1.32 & 17.81 \\
    000488 & 1.83 & 0.70 & 0.36 & 3.67 & 1.56 & 1.30 && 2.14 & 0.46 & 24.52 & 2.56 & 1.34 & 6.54 \\
    000539 & 2.01 & 0.59 & 4.51 & 6.92 & 1.93 & 10.51 && 2.17 & 0.45 & 1.41 & 2.73 & 1.36 & 43.14 \\
    000541 & 1.65 & 0.72 & 0.15 & 2.94 & 1.50 & 1.58 && 2.11 & 0.48 & 28.46 & 2.20 & 1.29 & 4.94 \\
    000550 & 2.15 & 0.62 & 2.68 & 6.93 & 1.84 & 1.54 && 2.21 & 0.46 & 6.67 & 3.48 & 1.41 & 23.27 \\
    000581 & 1.89 & 0.66 & 0.71 & 4.57 & 1.69 & 1.85 && 2.02 & 0.50 & 8.24 & 2.53 & 1.32 & 14.61 \\
    000625 & 2.12 & 0.65 & 1.99 & 5.90 & 1.74 & 0.80 && 2.19 & 0.46 & 12.02 & 3.40 & 1.41 & 14.11 \\
    000709 & 1.91 & 0.66 & 1.02 & 4.70 & 1.70 & 1.32 && 2.00 & 0.52 & 7.08 & 2.36 & 1.29 & 17.00 \\
    000720 & 1.51 & 0.66 & 9.53 & 2.77 & 1.59 & 22.33 && 1.90 & 0.55 & 8.26 & 1.83 & 1.22 & 29.66 \\
    000778 & 1.79 & 0.70 & 0.56 & 3.71 & 1.60 & 1.20 && 1.94 & 0.53 & 10.52 & 2.19 & 1.28 & 10.10 \\
    \hline \hline
 \end{tabular}
\end{table}

\begin{table}[htp]
 \centering
 \caption{\label{Tb:FitValuesPartialfilled} Values of fitted parameters,
$\alpha$, $\beta$ $q$, $\mu$ for the distribution of 23 stocks for
partial filled trades duration.}
 \medskip
 \centering
 \begin{tabular}{lccccccccccccc}
 \hline \hline
    Stock code  &  \multicolumn{6} {@{\extracolsep\fill}c}{MLE} & & \multicolumn{6}{@{\extracolsep\fill}c}{NLSE} \\
    \cline{2-7} \cline{9-14}
    &  $\alpha$ & $\beta$ & $\chi_w$ & $\mu$ & $q$ & $\chi_q$ & & $\alpha$ & $\beta$ & $\chi_w$ & $\mu$ & $q$ & $\chi_q$ \\
    \hline
    000001 & 1.68 & 0.70 & 0.56 & 3.17 & 1.55 & 4.84 && 1.78 & 0.58 & 1.83 & 2.21 & 1.29 & 11.34 \\
    000002 & 1.96 & 0.67 & 0.39 & 4.54 & 1.64 & 3.58 && 2.15 & 0.47 & 10.97 & 3.25 & 1.41 & 9.54 \\
    000009 & 1.85 & 0.66 & 0.58 & 4.30 & 1.67 & 4.45 && 2.02 & 0.49 & 7.28 & 2.95 & 1.40 & 11.74 \\
    000012 & 1.73 & 0.76 & 0.37 & 2.89 & 1.42 & 0.68 && 1.94 & 0.51 & 24.90 & 2.60 & 1.36 & 1.03 \\
    000016 & 1.48 & 0.80 & 0.11 & 2.09 & 1.33 & 0.49 && 1.78 & 0.54 & 23.38 & 2.10 & 1.31 & 0.48 \\
    000021 & 1.55 & 0.78 & 0.05 & 2.27 & 1.35 & 1.45 && 1.84 & 0.54 & 18.83 & 2.17 & 1.30 & 1.72 \\
    000024 & 1.67 & 0.77 & 0.68 & 2.67 & 1.40 & 0.42 && 1.88 & 0.52 & 26.20 & 2.38 & 1.33 & 0.58 \\
    000027 & 2.05 & 0.68 & 0.29 & 4.67 & 1.60 & 2.26 && 2.21 & 0.45 & 16.77 & 3.75 & 1.46 & 4.97 \\
    000063 & 1.81 & 0.75 & 0.65 & 3.17 & 1.45 & 0.34 && 1.98 & 0.51 & 24.39 & 2.66 & 1.35 & 1.06 \\
    000066 & 1.60 & 0.77 & 0.09 & 2.44 & 1.37 & 1.04 && 1.84 & 0.54 & 19.24 & 2.25 & 1.32 & 1.46 \\
    000088 & 1.79 & 0.74 & 1.93 & 3.30 & 1.49 & 0.18 && 1.94 & 0.51 & 27.17 & 2.62 & 1.36 & 0.83 \\
    000089 & 1.83 & 0.74 & 0.90 & 3.29 & 1.46 & 0.19 && 2.20 & 0.44 & 43.52 & 2.98 & 1.39 & 0.42 \\
    000406 & 1.54 & 0.79 & 0.11 & 2.26 & 1.35 & 1.01 && 2.04 & 0.47 & 39.77 & 2.35 & 1.34 & 0.84 \\
    000429 & 1.77 & 0.71 & 0.54 & 3.53 & 1.56 & 0.74 && 2.17 & 0.45 & 31.77 & 2.76 & 1.37 & 2.73 \\
    000488 & 1.60 & 0.81 & 1.09 & 2.38 & 1.34 & 0.24 && 2.05 & 0.47 & 53.93 & 2.43 & 1.35 & 0.31 \\
    000539 & 1.77 & 0.72 & 1.28 & 3.43 & 1.54 & 0.23 && 2.16 & 0.46 & 38.65 & 2.70 & 1.36 & 1.46 \\
    000541 & 1.59 & 0.79 & 1.24 & 2.45 & 1.38 & 0.33 && 2.05 & 0.47 & 50.28 & 2.35 & 1.33 & 0.28 \\
    000550 & 1.68 & 0.74 & 0.08 & 2.84 & 1.44 & 1.84 && 2.11 & 0.46 & 31.58 & 2.65 & 1.37 & 2.37 \\
    000581 & 1.83 & 0.74 & 1.42 & 3.40 & 1.49 & 0.34 && 2.02 & 0.50 & 28.21 & 2.80 & 1.37 & 0.96 \\
    000625 & 1.71 & 0.76 & 0.13 & 2.77 & 1.40 & 1.11 && 2.09 & 0.48 & 33.07 & 2.46 & 1.33 & 1.82 \\
    000709 & 1.81 & 0.70 & 0.15 & 3.71 & 1.58 & 1.72 && 1.96 & 0.52 & 11.35 & 2.61 & 1.35 & 6.20 \\
    000720 & 1.98 & 0.74 & 9.30 & 3.87 & 1.50 & 3.98 && 2.03 & 0.50 & 44.60 & 3.36 & 1.44 & 3.88 \\
    000778 & 1.57 & 0.78 & 0.32 & 2.40 & 1.38 & 0.60 && 1.73 & 0.57 & 14.80 & 2.20 & 1.32 & 0.98 \\
    \hline \hline
 \end{tabular}
\end{table}

\end{document}